\def\'#1{\ifx#1i{\accent"13\i}\else{\accent"13#1}\fi}

\def\alamenos#1{$^{-#1}$}

\def\be{\begin{equation}}
\def\diezalamenos#1{10$^{-#1}$}
\def\diezala#1{10$^{#1}$}
\def\ee{\end{equation}}
\def\gc{GC}
\def\grados{$^\circ$}

\def\lsd{LSD}
\def\Ncore{$N_{\rm 0,core}$}
\def\ncerocore{{$n_{0\ {\rm core}}$}}
\def\ncerofit{{$n_{0\ {\rm fit}}$}}
\def\ncore{$n_{\rm 0\ core}$}
\def\nfit{$n_{\rm 0\ fit}$}
\def\por{$\times$}
\def\Rcore{$R_{\rm core}$}

\def\ssd{SSD}
\def\tcero{$t_0$}
\def\tuno{$t_1$}
\def\Tcore{$T_{\rm core}$}
\def\Tfit{$T_{\rm fit}$}

\def\ximax{{$\xi_{\rm max}$}}

\def\xy{{$x$-$y$}}
\def\xz{{$x$-$z$}}
\def\yz{{$y$-$z$}}

\documentclass[preprint]{aastex}

\usepackage{emulateapj5}
\usepackage{graphicx}

\begin{document}

\title{Dynamic cores in hydrostatic disguise}

\author{Javier Ballesteros-Paredes$^{1,2}$, Ralf S.\ Klessen$^{3,4}$, and
Enrique V\'azquez-Semadeni$^2$}

\affil{$^1$Department of Astrophysics, American Museum of Natural
History\\ Central Park West at 79th Street, New York, NY, 10024-5192,
U.S.A.;} 

\affil{$^2$Instituto de Astronom\'ia, UNAM \\ 
{\tt j.ballesteros@astrosmo.unam.mx}; {\tt e.vazquez@astrosmo.unam.mx}}

\affil{$^3$UCO/Lick Observatory, University of California, Santa Cruz, CA
95064, U.S.A.;\\ } 

\affil{$^4$Astrophysikalisches Institut Potsdam, An der Sternwarte 16,
  14482 Potsdam, Germany; \\
{\tt rklessen@aip.de}}

\slugcomment{Draft date: \today}

\lefthead{Ballesteros-Paredes et al.}

\righthead{Dynamic cores in hydrostatic disguise}

\begin{abstract}

%numero total de cores rechazados=          16
%numero total de cores con 3 proyecciones BE=          86
%numero total de cores con 2 proyecciones BE=          19
%numero total de cores con 1 projeccion BE           6

We discuss the column density profiles of ``cores'' in
three-dimensional SPH numerical simulations of turbulent molecular
clouds. The SPH scheme allows us to perform a high spatial resolution
analysis of the density maxima (cores) at 
scales between $\sim$~0.003 and 0.3~pc. We analyze simulations in
three different physical conditions: large scale driving (LSD), small scale
driving (SSD), and random Gaussian initial conditions without driving (GC);
each one at two different timesteps: just before self-gravity is
turned-on (\tcero), and when gravity has been operating such that 5\%
of the total mass in the box has been accretted into cores (\tuno). 
For this dataset, we perform
Bonnor-Ebert fits to the column density profiles of cores found by
a clump-finding algorithm. We
find that, for the particular fitting procedure we use, 65\%\ of the
cores can be 
matched to Bonnor-Ebert (BE) profiles, and of these, 47\%
correspond to {\it stable} equilibrium configurations with \ximax $< 6.5$,
even though the cores analyzed in the simulations are not 
in equilibrium, but instead are dynamically evolving. The
temperatures obtained with the fitting procedure vary between 5 and
60~K (in spite of the simulations being isothermal, with $T=$ 11.3~K),
with the peak of the distribution being at $T=$ 11~K, and most clumps
having fitted temperatures between 5 and 30~K.
Central densities obtained with the BE fit tend to be smaller than the
actual central densities of the cores. We also find that for the LSD
and GC cases, there are more BE-like cores at \tcero\ 
than at \tuno\ with $\xi_{\rm max} \le 20$, while in the case of SSD,
there are more such cores at \tuno\ than at \tcero. We interpret this as 
a consequence of the stronger turbulence present in the cores of run
\ssd, which prevents good BE fits in the absence of gravity, and delays
collapse in its presence.
Finally, in some cases  we find substantial superposition effects when we
analyze the projection of the density structures, even though the
scales over which we project are small ($\lesssim 0.18$~pc). As a
consequence, different projections of the same core may give very different
values of the BE fits. Finally, we briefly discuss recent results
claiming that Bok globule B68 is in hydrostatic equilibrium, stressing
that they imply that this core is unstable by a wide margin.
We conclude that fitting BE profiles to
observed cores is not an unambiguous test of hydrostatic
equilibrium, and that fit-estimated parameters like mass, central
density, density 
contrast, temperature, or radial profile of the BE sphere may differ
significantly from the actual values in the cores.  

\end{abstract}


\section{Introduction}\label{intro}

\def\BP{Ballesteros-Paredes}
\def\VS{V\'azquez-Semadeni}

Stars and planets form in dense cores within molecular clouds. In the
case of low-mass star forming regions, it has traditionally been
thought that these cores are quasi-static equilibrium configurations
supported against gravitational collapse by a combination of magnetic
and thermal pressures (see, e.g.~Shu, Adams \& Lizano 1987). When considering
purely thermal support, stable hydrostatic solutions of the self-gravitating
fluid equations with finite central densities and sizes exist,
provided the sphere is confined by an external pressure $P_{\rm ext}$
acting on the surface. These solutions are known as Bonnor-Ebert (BE)
spheres \citep{Ebert55, Bonnor56}. Recently, Alves, Lada \&\
Lada(2001) have determined with small error bars the column density
profile of B68, a core embedded in an HII region. They have also
successfully fitted a BE profile. Similar studies
\citep{Johnstone_etal00, Harvey_etal01, Evans_etal01} have attempted
to fit BE profiles to the column density profiles of selected molecular
cloud cores.  

%In particular, the study by \citet{Alves_etal01} has used
%state-of-the-art star counting techniques in deep visible and
%near-infrared images (B, V, I and K bands) taken with ESO's Very Large
%and New Technology telescopes. These authors chose to angle-average
%the data when producing the column density profiles, and that the
%effective resolution over the core area was only $\sim 30\times 30$
%pixels, since only less than a thousand stars lie behind the core.

However, the picture of isothermal cores in hydrostatic equilibrium
may be in conflict with the fact that molecular clouds are turbulent.
It appears difficult that quasistatic equilibrium
structures may appear and survive in isothermal, supersonic, highly
compressible turbulent flows, where density fluctuations are in
general transient, and have further substructure in a self-similar
hierarchy \citep{Scalo85, Houlahan_Scalo90, Falgarone91,
Houlahan_Scalo92, VS94, VS_etal00, Williams_etal00, Chappell_Scalo01,
%RSK:modified
Klessen01, MacLow_Klessen03},
which may possibly end at small enough scales that the
turbulent velocity fluctuations are no longer supersonic, and cannot
produce further turbulent fragmentation\footnote{Note that by small
enough scale we just mean a typical average scale at which the
velocity dispersion equals the sound speed, but around which there
may be a large scatter in the sizes of individual clumps satisfying
this condition.} (V\'azquez-Semadeni,
Ballesteros-Paredes \& Klessen 2003a). Nevertheless, in the turbulent
case, there is no reason why these structureless density fluctuations
should be hydrostatic, as they have formed from larger-scale turbulent
compressions within a uniform-temperature medium. It is more likely
that they should either
re-expand or proceed to gravitational fragmentation and collapse,
since the probability of them reaching precise  
balance between self-gravity and their internal pressure is
vanishingly small (Tohline, Bodenheimer \& Christodoulou 1987;
Taylor, Morata \& Williams 1996; Ballesteros-Paredes,
V\'azquez-Semadeni \& Scalo 1999; V\'azquez-Semadeni, Shadmehri \&
Ballesteros-Paredes 2003b).  Note that the condition for
constructing a BE sphere, 
namely the availability of a hotter, more tenuous confining medium
is in general not realized between molecular clouds and their embedded cores,
as they are both at roughly the same temperature.
%Also, note that the latter authors have
%remarked that in fact, 
%BE equilibrium spheres cannot exist if the ``external'' confining
%medium is at the same temperature as the sphere itself, because in
%this case there is no real the pressure decreases outwards linearly
%with the density. 

The transient 
character of most turbulent fluctuations and the eventual induction of
collapse on one of them in a gravitationally unbound medium was shown by
V\'azquez-Semadeni, Passot \& Pouquet (1996). Moreover, simulations of
turbulent, globally 
gravitationally bound clouds consistently show the collapse of local
peaks, but never the formation of hydrostatic cores, either in the
purely hydrodynamic case (L\'eorat, Passot \& Pouquet 1990; Klessen,
Heitsch \& Mac Low 2000) or in the magneto-hydrodynamic (MHD) case
\citep{HMK01}. One exception is the 
case of MHD simulations in boxes with subcritical mass-to-magnetic
flux ratios (Ostriker, Stone \& Gammie 1999), which evolve towards flattened
sheets that cannot collapse in the direction perpendicular to the
field. However, being performed in closed boxes, these simulations
cannot represent the fact that more mass can continue to accrete along
field lines until a core becomes supercritical (Hartmann,
Ballesteros-Paredes \&\ Bergin 2001). There also exist both
observational evidence \citep{Crutcher99, Andre_etal00, Bourke_etal01,
HBB01} and
theoretical arguments 
% RSK: modified \citep{Nakano98, HBB01} 
(Nakano 1998; Hartmann et al.\ 2001; as summarized by Mac~Low \&
Klessen 2003)
showing that the
overwhelming majority of cores must be supercritical. Moreover, the
long lifetimes that quasi-static cores would have are difficult to
reconcile with observational statistics of cloud cores
\citep{Taylor_etal96, Lee_Myers99, Visser_etal02}, and with the
suggestion of short molecular cloud 
formation time scales ($\sim$ a few Myr), based on the observed lack
of post-T-Tauri stars in Taurus (Herbig, 1978; Ballesteros-Paredes,
Hartmann \& V\'azquez-Semadeni 1999; Hartmann 2002).

It is important
to mention that the preceding discussion is not in contradiction
with the fact that stars are objects in hydrostatic equilibrium
within a turbulent medium, since they do not correspond to
quasi-isothermal flows. In stars, energy is trapped since the
opacity has increased, and the cooling time is about \diezala{10}
times the free-fall time. This is also consistent with the fact that
the formation of a stable hydrostatic structure requires a
polytropic exponent $\gamma > 4/3$, where $P\propto\rho^\gamma$
\citep{Chandrasekhar}. On the other hand, 
molecular clouds and their cores are not able to trap energy
effectively because they are approximately isothermal
\citep{Goldsmith_Langer78, Goldsmith88, Scalo_etal98, BVS99,
Spaans_Silk00}. Thus, they are not expected to be able to reach a
hydrostatic configuration if they 
are formed by a turbulent compression (Ballesteros-Paredes, Hartmann
\& V\'azquez-Semadeni 1999; V\'azquez-Semadeni, et. al 2003b). This
conclusion may hold even when magnetic fields are considered because,
as mentioned above, the existence of subcritical cores is being
questioned, and supercritical configurations are qualitatively
equivalent to non-magnetic ones. Finally, even is magnetically
subcritical cores do exist in clouds, their flattened configurations are very
different from those of the BE sphere, and no reason exists for them
to be well fitted by a BE profile.
 
In view of the above, it is clear that the recent attempts to fit 
hydrostatic BE profiles to observed molecular cloud cores appear
to be in serious contradiction with the theoretical results suggesting
that non-magnetic hydrostatic structures are unlikely to form in a turbulent
medium and that the only hydrostatic magnetic configurations ought to
be flattened. In this paper, we show that the apparent discrepancy can be
resolved because fitting a BE-type profile to an observed cloud core
is not an unambiguous test of it being in hydrostatic equilibrium. To
show this, we take some of the cores (which are not in
hydrostatic equilibrium) in the numerical simulations of Klessen,
Burkert \& Bate (1998), \citet{Klessen_Burkert00},
\citet{Klessen_Burkert01}, and \citet{KHM00}, and apply to them a
fitting procedure  similar to that used by \citet{Alves_etal01},
showing that reasonable BE column-density profile fits can be made on
these non hydrostatic structures as well. 

The plan of the paper is as follows. In \S\ref{simulations} we summarize
the main characteristics of the simulations analyzed. In \S\ref{analisis}
we explain the numerical technique in fitting BE column density
profiles, and in \S\ref{sec:results} we present the results of the
fits. In \S\ref{discussion} we discuss the results, and present the
main conclusions.

\section{Numerical Simulations}\label{simulations}

A prerequisite for adequately describing the inner structure of cores
in models of turbulent molecular clouds is  to use a numerical
technique that is able to resolve high density contrasts, at arbitrary
locations within the cloud. The method of choice is smoothed particle
hydrodynamics (SPH), which is a particle-based Lagrangian scheme to
solve the equations of hydrodynamics. Flow properties are obtained by
averaging over an appropriate subset of SPH particles, and high
resolution is achieved where needed by increasing the particle
concentration and reducing the averaging volume \citep{Benz90,
Monaghan92}. 

For the current investigation we re-analyze numerical models of
turbulent molecular cloud evolution first presented by \citet{KBB98,
Klessen_Burkert00, Klessen_Burkert01} and \citet{KHM00} using 205,379
particles. These calculations solve the equations for 
a self-gravitating isothermal ideal gas in cubic volumes with periodic
boundary conditions, and are imagined to be located within larger,
roughly self-similar clouds. For the density range of interest
($10^2\,{\rm cm}^{-3} \le n({\rm H}_2) \le 10^7\,{\rm cm}^{-3}$)
interstellar molecular gas is 
seen at temperatures close to $10\,$K and the isothermal equation of
state is approximately valid \citep{Goldsmith_Langer78,
Goldsmith88,Scalo_etal98, BVS99, Spaans_Silk00}.  Under these
conditions the dynamical 
behavior of the gas is scale-free and depends only on the ratio of
internal and turbulent kinetic energy to gravitational energy. The
numerical simulations are thus performed in dimensionless units and
need to be rescaled to obtain physical quantities. In the
simulations, once the density in a region exceeds a density contrast
of $\sim 10^4$, a sink particle is 
created, with the same mass of the region where it
formed, but with a fixed radius of the order of the Jeans
length at the threshold density \citep{Bate_etal95}. The
internal structure of sink particles is not resolved.

To facilitate a direct comparison with the observations we decompose
the gas distribution in each numerical model into individual clumps
using a 3-dimensional clump-finding algorithm (see Appendix A in
Klessen \& Burkert 2000) and select the first 21 gas clumps without
sink particles (as their analysis would be meaningless)  determined by
the  clump-finding algorithm.  
%RSK: double ``as'' removed

Once we have the individual clumps, we define the centers
of control subregions (with volume $(1/10)^3$ of the full simulation
cube) as the location of the density peak. The SPH density
distribution within a control subregion is assigned 
onto a $128^3$ grid, and integration along the principal axes produces
three column-density maps -- one for each direction of projection.
Note that the location of 
the density peak needs not correspond to the maximum of the projected column
density. 

To compare our maps with observed stellar extinction maps of dark
globules, we adopt a physical scaling such that, for a typical
protostellar core in the simulation, the resulting column density
map and pixel sizes roughly correspond to the observations of the Bok
globule B68 by \citet{Alves_etal01}. Our maps then have a physical
size of $0.18\,$pc$\times0.18\,$pc and column densities $N({\rm H}_2)$
ranging from about $10^{20}\,$cm$^{-2}$ for the background value to a
few times $10^{22}\,$cm$^{-2}$ at the core centers. For details on the
scaling procedure the reader is referred to \citet{Klessen_Burkert00}
and \citet{KHM00}.

We analyze three different numerical models. These constitute the
extreme ends of the possible range of molecular cloud dynamics. In the
first two models, the molecular gas is subject to strong supersonic
turbulence. In the first (hereafter \lsd),
turbulence is driven on large scales, with
wavelength $\lambda \approx 1/2$ of the computational box. In the
second model (denoted \ssd), energy is injected on small scales, with
$\lambda\approx 1/8$ of the computational box. With a box size of
1.54 pc per side, these correspond to driving scales of $\sim
0.77\,$pc and $0.19\,$pc, respectively.  For each model we consider
two different evolutionary stages. First, a stage ($=t_0$) of fully
developed supersonic turbulence without self-gravity,
and second, one snapshot after self-gravity has been switched on and
gravitational contraction has led to the formation of collapsed cores
accumulating roughly 5\% of the total mass ($=t_1$). These two models
are complemented by one simulation without driving, and where the gas
is allowed to collapse freely from a field of random Gaussian density
fluctuations (hereafter \gc) under the influence of
self-gravity. Again, two times are considered, $t=t_0$, now
corresponding to the evolutionary phase just before the first
collapsed core occurs,  and $t=t_1$,
when $\sim 5$\% of the mass is in collapsed objects. For a more
detailed discussion of the dynamical evolution the models and the
implications for star formation, see \citet{Klessen01}. The properties
relevant for the current investigation are summarized in Table
\ref{tabla:models}. 

%%%%%%%%%%%%%%%%%%%%%%%%%%%%%%%%%%%%%%%%%%%%%%%%%%%%%%%%%%%%%%%%%%%%%%%%%%%%%%
% NEW (R.S.K.)
%%%%%%%%%%%%%%%%%%%%%%%%%%%%%%%%%%%%%%%%%%%%%%%%%%%%%%%%%%%%%%%%%%%%%%%%%%%%%%

\section{Analysis of the Simulations}\label{analisis} 

\subsection{Bonnor-Ebert Models}\label{BE}

In the classical analysis of \citet{Ebert55} and \citet{Bonnor56}, the
equation that describes a self-gravitating isothermal sphere in
hydrostatic equilibrium is the modified Lane-Emden equation: 

\begin{equation}
{d\over d\xi}\biggl(\xi^2 {d\Psi\over d\xi}\biggr) = \xi^2\ e^{-\Psi} 
\label{lane}
\end{equation}
where $\xi=(r/c_s)\sqrt{4\pi G\rho_{\rm 0 core}}$ is a non-dimensional
radial variable, $r$ is the radial coordinate, $c_s = \sqrt{kT_{\rm
core}/m}$ is the thermal velocity at temperature $T_{\rm core}$, $k$
is the Boltzmann constant, $G$ is the gravitational constant,
$\rho_{\rm 0core}=m_p n_{\rm 0core}$ is the central mass density, $m_p$
is the mean mass per particle, $n_{\rm 0 core}$ is the central number
density of 
particles, and $\Psi=-\ln{(\rho/\rho_{\rm 0 core})}$.  The solutions
of eq. (\ref{lane}) are required to be finite at the center of the
core, which implies that they must satisfy the boundary conditions
$\Psi(0)=0 \label{ci1a}$ and ${d\Psi/d\xi}(0)=0.$ 

Equation (\ref{lane}) can be solved by the change of variables
$y_0=\xi^2 \eta$, $y_1=\Psi$ (with $\eta=d\Psi/d\xi$), which splits up
the equation into two first-order equations:  

\begin{eqnarray}
{dy_1\over d\xi}&=&{y_0\over \xi^2} \label{tosolve1}\\
{dy_0\over d\xi}&=&\xi^2\exp{(-y_1)}\label{tosolve2},
\end{eqnarray}
with the initial conditions $y_0(0)=0$, and $y_1(0)=0$. This system
has a unique solution out to the nondimensional parameter 

\be
\xi_{\rm max} = {R_{\rm core}\over c_s}\sqrt{4\pi G\rho_{\rm 0 core}},
\label{ximax_eq}
\ee
where \ximax\ is the value of $\xi$ at the outer boundary $R_{\rm
core}$, beyond which the density is negligible, but the temperature is
assumed to be sufficiently high as to maintain pressure
equilibrium at the boundary.  

Here we stress two points. First, a particular solution is
defined by the radius, temperature and central density
of the core, \Rcore, \Tcore, and $\rho_{\rm 0 core}$;
and second, if the solution has $\xi_{\rm max} >$ 6.5 (i.e., if
the density contrast between the center and the edge of the BE sphere
is larger than 14.3), the configuration is in an unstable
equilibrium. We will get back to these points in next section.

\subsection{Numerical Technique}\label{numerics:BE}

\subsubsection{Construction of column density maps and radial
profiles}\label{maps_profiles}  

In order to mimic the observational procedure, we assume that the
center of the BE sphere is located at the position of the column
density maximum, $(x_0,y_0)$. Then, we  slice the column density map
into 36 angular sectors of width 10\grados,  centered on
$(x_0,y_0)$. For each sector we construct the 
radial column density profile. If in a particular sector more than
one measurement falls into the same radial bin, we take the arithmetic
average. This procedure gives a column density profile in which the 
values of the abscissas (the radii) are unevenly spaced. We thus
redefine the profile 
as an interpolation of the latter at homogeneously
distributed radii. 
Finally, we define the angle-averaged radial column density
as the average profile over the 36 slices. We also define 
``dispersion'' profiles as the mean profile $\pm$ the standard deviation
of the 36 profiles. We generically refer to
both of them as the ``dispersion profile''.

Our cubic subregions typically contain several thousand
particles. For the inner parts (i.e. the central regions of the
considered clumps), the subregion resolution of $128^3$ cells roughly
reflects the effective spatial resolution 
of the SPH method in these high-density regions. Towards the outer
parts, where the density is lower, the smoothing volume of an SPH
particle may exceed the cell size which we adopt, so that
in the outer low-density wings of the cores we do not fully
resolve the small-scale structure. However, this does
not influence the BE fitting procedure significantly, because we use
logarithmic bins in the radial direction for constructing the radial
profiles, implying that we have larger 
stepsizes in the low density regions corresponding to the lower spatial
resolution there. We expect that the additional smoothing in the wings
leads to a slight overestimate of the ``true" density (i.e. if
we had infinite mass and spatial resolution), because we have a
positive gradient inwards. However, there being one or more decades in
column density below the peak, we expect the effect on the
overall result of the BE fit to be negligible.

\subsubsection{Fitting procedure}\label{fitting:sec}

We calculate the solutions of the modified Lane-Emden equation by
solving equations (\ref{tosolve1}) and (\ref{tosolve2}) using an
ordinary fourth order Runge-Kutta method. Similarly to the criterion
used by Alves et al.\ (2001), who
fitted the observational data only out to the radius where the column
density profile started to deviate significantly from a BE-like profile
(roughly 100 arcsec in their fig.\ 2; J.\ Alves, private communication), 
in this paper we
choose the value of outer radius of the core ($R_{\rm 
core}$) by inspection of the column density maps, selecting either the
radius at which noticeable kinks start to appear in the average column
density profile, or else that at which the profile begins to
clearly deviate from a BE one.
%(often, the profiles at intermediate distances from the 
%peaks approach power laws). 
Subsequently, we calculate a
set of BE spheres by varying both the central density and the
temperature (see below), and we define the BE profile that ``best
matches'' the core as the one that minimizes the function: 

\be
\langle \chi^2 \rangle={1\over m}\sum_i{\biggl[\log{(N_{i,{\rm
BE}})}-\log{(N_{i,{\rm core}})}\biggr]^2},  
\label{goodness}
\ee
where $N_{{\rm BE},i}$ and $N_{{\rm core},i}$ are, respectively,
the column density of the fitted Bonnor-Ebert
sphere, and of the actual column density of the
core at the $i$-th radius, and $m$ is the number of points used to
construct the radial profile. The sum extends over all points at
which $N_{{\rm BE}}$ and $N_{{\rm core}}$ are evaluated (i.e., $i=1,
\dots, m$). We have chosen to minimize the difference of the logarithms
in order to give the same weight to the central and outer parts of the
cores, even though they have very disparate typical column densities.

The ranges of central density and temperature values used to find the
best BE fit are, for $\rho_{\rm 0 core}$ in eq (\ref{ximax_eq}),
from 0.1 to 10 times the actual value of the central density, with
linear increments of size 0.1 below this density, and of size 1 above it.
For the temperature, we actually vary the value of 
\ximax, from 1 to 20 with increments of 0.1, which is equivalent to varying
the sound speed, and thus the temperature (see eq.~[\ref{ximax_eq}]). 

A few comments regarding the choice of method are in order. Our
method follows as closely as possible those used in recent
observational works (e.g., Johnstone et al.\ 2000; Evans et al.\ 2001;
Alves et al.\ 2001; Harvey et
al.\ 2001), although in general not much detail is given in those
papers. Other choices for the fitting procedure are  
clearly possible. In fact, when analyzing the simulations there is more
freedom in choosing the procedure, because {\it we have
much more information about the physical fields than is available
observationally.} Observational data generally cover only the core
itself, and extend only out to where the data become noisy.
In the simulations, one has information on the density field out to
distances up to two orders of magnitude larger than the
core, and moreover, there
is no instrument noise. This means that {\it there is no clear
boundary to the cores}. Thus, the choice of the truncation radius
becomes more arbitrary. Our criterion, based mostly on the shapes of the 
profiles, is probably as arbitrary as one based on where the data become 
too noisy. We could have instead chosen to truncate our fits where the
column density contrast reaches a certain threshold, but this would
probably bias our results towards BE fits with a certain \ximax\ (as
is probably the case in some of the 
observational papers). Alternatively, one could fix the central density and
temperature, and then use $R_{\rm core}$ and the external pressure as
the fitting parameters, but this is not what is done observationally.
Keeping in mind our goal of showing that those works do not provide
unambiguous evidence of the hydrostaticity of the cores, we consider
that our conventions are adequate.

\section{Results}\label{sec:results}

\subsection{Statistical Analysis}\label{sec:statistical}

We have analyzed the first 21 cores found by the clump finding
algorithm, at two different timesteps of each one of 
the three simulations. Since we have projected each core  on its 3
directions, we have analyzed a total of 378 column density profiles. We
stress again that none of these cores are in hydrostatic
equilibrium. Instead, they either get dispersed or collapse (see
\S\ref{sec:nonhydro}). 

Table~\ref{tabla:results} lists the results of the analysis for one
core in each physical situation, chosen to illustrate the variety of
situations that can occur.
%\footnote{The interested reader can
%consult the whole table at {\tt http://www.astrosmo.unam.mx/\~\
%j.ballesteros/BE/table.ps} {(\bf enro and ralf, please check that you
%can get the table through the WEB)}{\bf enro suggests to do it also
%in the electronic version of the 
%paper} ---- \{J+E: I cannot read the table. I can get to the
%directory, but cannot see a file ``table.ps''.\} --- \{J+E: Yes, we
%should give the full table in the electronic version of the paper.\}
%}. 
Column 1 gives the name of the
run (see \S\ref{simulations} for its physical parameters.) Columns 2,
3 and 4 denote the time at which the run was analyzed, the order number of
the core, and the projection plane, respectively. Columns 5 and 6
denote the central density of the core (\ncerocore) and the central
density of the fitted BE sphere (\ncerofit), respectively. Columns 7
and 8 respectively give the temperature (\Tfit) and the value of the
nondimensional  radius $\xi_{\rm max}$ for the best core fit. These values
have been obtained using a radius (\Rcore) listed in column 9. Column
10 gives the value of the rms error, given by eq. (\ref{goodness}),
between the fit and the actual angle-averaged column density profile,
$\langle \chi_1^2 
\rangle^{1/2}$. Similarly, in column 11 we give the rms separation
between the ``dispersion'' profile and the angle-averaged one, $\langle
\chi_2^2 \rangle^{1/2}$.

%numero total de cores rechazados=          32
%numero total de cores con 3 proyecciones BE=          37 = 111
%numero total de cores con 2 proyecciones BE=          32 = 64
%numero total de cores con 1 projeccion BE          25    = 25

\begin{figure} 
\begin{center}
\includegraphics[width=0.7\textwidth]{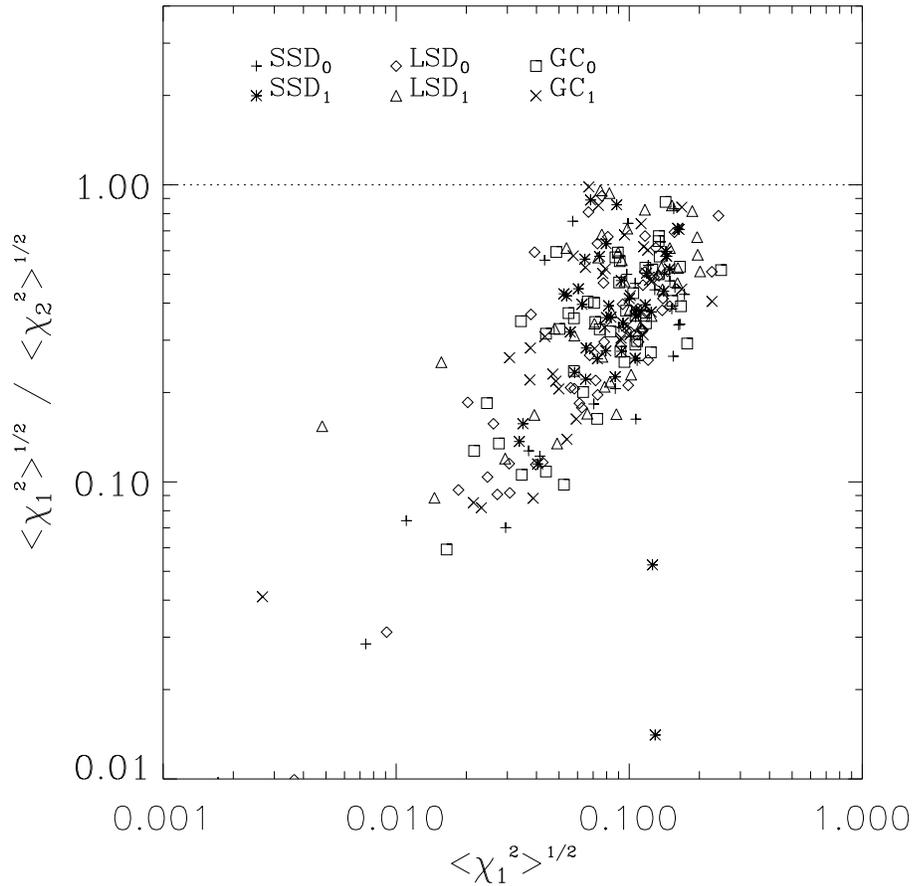}
\end{center}
\caption{Ratio of $\langle \chi_1^2 \rangle^{1/2} / \langle \chi_2^2
\rangle^{1/2}$ vs. $\langle \chi_1^2 \rangle^{1/2}$ for the cores
with  good BE fit. Since the rms distance between a good fit
and the actual profile must be smaller than the rms distance between the
dispersion curve and the actual profile, the points fall below the
dotted line. 
\label{chi_chi}}
\end{figure}

 From the 378 projections, 
we reject those that have fits whose rms distance to the
column density profile (\Ncore) is larger than the rms distance of the
dispersion curves to \Ncore, i.e., if $\langle \chi_1^2 \rangle^{1/2} >
\langle \chi_2^2 \rangle^{1/2}$. Thus, only 65.08\%\ of the 378
projections are well described by 
a BE profile, for which the values of the ratio $\langle \chi_1^2
\rangle^{1/2} / \langle \chi_2^2 \rangle^{1/2}$ are plotted against
$\langle \chi_1^2 \rangle^{1/2}$ in Fig.~\ref{chi_chi}. From this
figure, we note that the bulk of the fitted profiles are roughly a factor
of 2 or 3 closer to the mean column density profile than the dispersion
curves.  

\begin{figure} 
\begin{center}
\includegraphics[width=0.7\textwidth]{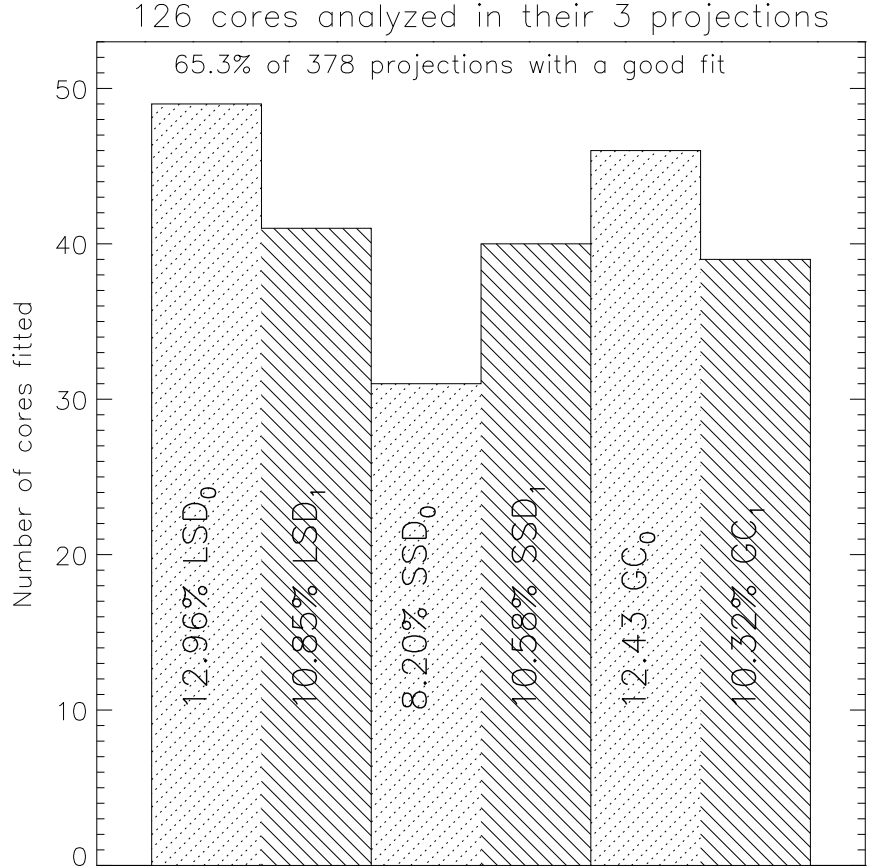}
\end{center}
\caption{Number of cores with good BE fit. \lsd\ is represented by the
first two columns. \ssd\ is represented by the middle two, and \gc\
is represented by the last two columns. Dotted lines at 45 degrees
represent cores at $t=t_0$, and solid lines at
-45 degrees represent cores evaluated at $t=t_1$. 
\label{histogram}}
\end{figure}

In Fig.~\ref{histogram} we show the number of projections that have
good BE fits, for all run types and times considered. We note that,
for runs \gc\ and \lsd, there are fewer cores with BE-like fits at \tuno\
than at \tcero. In contrast, for run \ssd, there are substantially more
BE-like cores at \tuno\ than at \tcero. These results suggest that the
structure of 
the cores depends sensitively on the global parameters and initial
conditions of the flow. Indeed, at $t=$\tcero\, the cores in runs \lsd\
and \ssd\ are due exclusively to advection, the ones in run \ssd\ having 
more substructure \citep{MacLow_Ossenkopf00, BM02}, because the
smaller-scale driving implies that there is more turbulent energy at
small scales in this run than in \lsd. Run \gc, on the
other hand, was started with an already relatively smooth initial
density distribution and furthermore has been subject to the action of
self-gravity from the start, so it too, as run \lsd, has little
substructure within the cores. Thus, a BE profile (with $\xi_{\rm max} \le 
20$) is a better fit for
the smoother cores of runs \gc\ and \lsd\ than for the more irregular
cores in run \ssd\ at $t=$\tcero.

%may be understood as
%follows: \gc\ models do not contain initial substructure, since they
%are random Gaussian density fluctuations with most of the power at
%large scales ($P(k)\propto k^{-2}$, with $P$ the power spectrum and
%$k$ the wavenumber, see\citet{Klessen_Burkert00}){\bf (Ralf, plese
%check this)}. In the case of \lsdcero, where there is a random forcing
%at large scales without gravity, it is found that the density
%structures do not have substantial substructure below 0.1 pc
%\citep{MacLow_Ossenkopf00, BM02}. Although the origin of the lack of
%substructure is different in both models, this lack allows for more
%roundish cores which when are projected and circle-averaged, they may
%mimic a BE profile. Finally, when a simulation without gravity is
%forced at small scales, as is the case of \ssdcero, it is found that
%there is substantial substructure at small scales
%\citep{MacLow_Ossenkopf00, BM02}, and the BE fit becomes more
%difficult. 

At \tuno, instead, gravity has already been acting for some time on all
runs, and has produced both smoothing and collapse of the cores. In this 
case, the originally smoother cores in runs \lsd\ and \gc\ have 
%RSK: removed  ``probably''
already reached advanced stages of collapse, and fitting them would
%RSK: removed ``probably''
require larger values of $\xi_{\rm max}$ than we have considered
here, while the cores in run \ssd, which started out more turbulent, had 
to first overcome that turbulence (either by dissipating it or by
accreting more mass) and only then could start collapsing. Thus, they
%RSK: modified: are probably 
appear to be in an earlier collapse phase with more of them in
the range $1 \le \xi_{\rm max} \le 20$. Note that, however, the above
results do not imply that less mass has been accreted into collapsed
objects in run \ssd\ compared to the other two, since \tuno\ 
is defined as the time at which this mass is 5\% of the total. Instead,
it is simply a reflection of the fact that collapse proceeds at a higher 
{\it rate} in runs \lsd\ and \gc\ \citep{KHM00}.

%This situation is reversed at \tuno, once the gravity has played a
%role. In this case, the original cores in \gc\ and \lsd\ have had time
%to collapse and become compact structures with large density contrast,
%and a BE fit with values of \ximax\ smaller than 20 becomes more
%difficult to find. Another posibility is that collapsed objects just
%do not resemble a projected BE sphere. As a consequence, we obtain a
%lower number of cores with a good BE profile. Instead, for \ssd\ at
%\tuno, the collapse has been slower than for \lsd, the cores had have
%time to develope central condensations, and the internal substructure
%may be somehow errased by the action of gravity, allowing for a more
%BE-like profiles. 

\begin{figure} 
\plotone{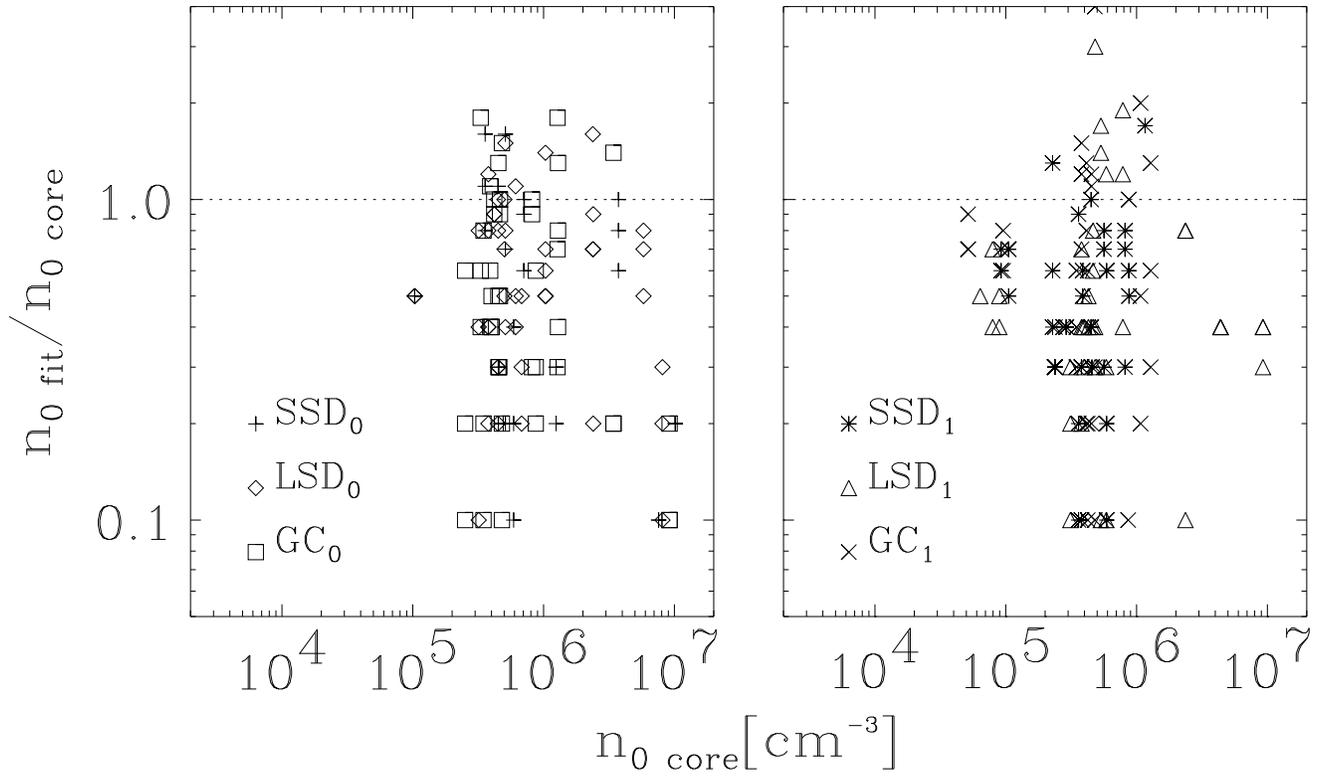}
\caption{Ratio \ncore/\nfit\ as a function of \ncore. Note that there
is a tendency of \nfit to be smaller than \ncore, but no a clear
trend is found regarding the time under analysis.
\label{densities}}
\end{figure}

We now turn to the analysis of the physical parameters as obtained
from the fitting procedure. Fig.~\ref{densities} shows the ratio of
fitted-to-actual central density, \ncerofit/\ncerocore, as a function
of \ncerocore\ at times \tcero\ (left panel) and \tuno\ (right
panel). We find that the central density obtained with the BE fit
tends to be smaller than the actual density of the core. 
%
%
%the actual central density of the core, \ncerocore, for the times
%\tcero\ (left panel) and \tuno\ (right panel). Although it can be
%expected that the central core density \ncerocore\ should be larger at
%\tuno\ than at \tcero, this is not reflected when comparing the left
%and the right panels. This result is due to the fact that we have
%removed from the analysis the most collapsed cores, which include sink
%particles, as we mentioned before. A more important result 
%is the fact that the central density obtained with the
%BE fit tends to be smaller than the actual central density of the
%core. 
%This suggests that densities obtained with the BE fit procedure
%systematically fall short of the real values in the
%core.

\begin{figure} 
\plotone{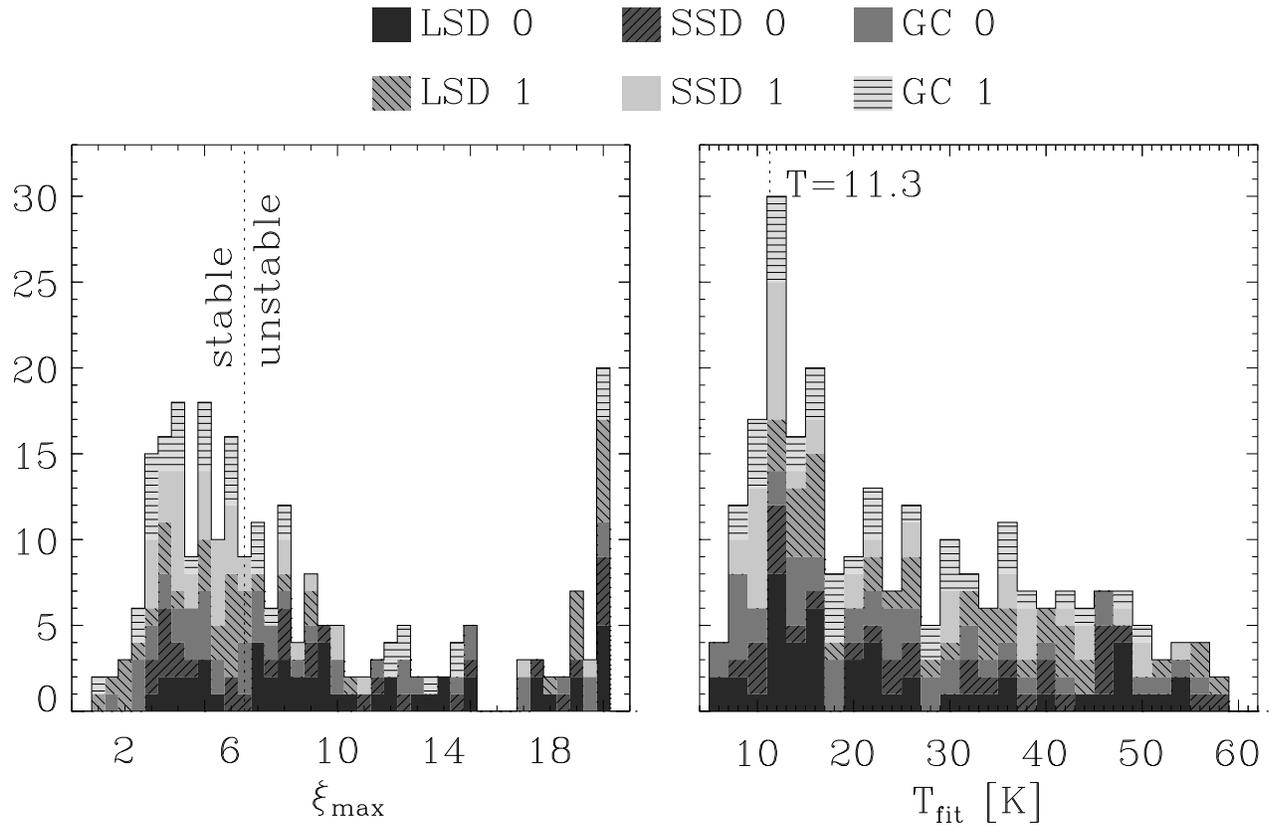}
\caption{Histograms of (a) \ximax, and (b) \Tfit. The
contribution of the different models are denoted by the different
grayscale tone. Regarding \ximax, note that approximately 1/2 of the
cores exhibit  
a stable BE profile (47.15\%). Regarding the temperature, the
distribution is broad, ranging from 5 to 60. Note also that the
simulations are isothermal, implying that the 
fitted temperature can not correspond to the actual temperature of the
core. 
\label{histogramas}}
\end{figure}

In Fig.'s~\ref{histogramas}a and b, we show the distribution of the
fit
parameters \ximax, and \Tfit. The contribution of 
each one of the six physical situations is 
represented with a different gray-scale tone.  In the first histogram,
the vertical dotted line at \ximax~$=$~6.5 denotes the critical value above
which a BE density profile corresponds to an unstable
equilibrium. We note that roughly half (47.2\%) 
of the fits have \ximax$\le 6.5$, making the cores to appear
as stable configurations. In
Fig.~\ref{histogramas}b the histogram exhibits a broad distribution of
fit temperatures, ranging between 5~K and 60~K, although most of
them have temperatures between 5 and 30~K. Recall that the numerical
simulations are isothermal, with the scaling taken such that $T =
11.3$~K. Thus, we see that the 
temperature derived from the BE fit procedure does not recover the
actual temperature of the core very well. Finally, in
Fig.~\ref{histogramas}c (right), we show that most of the fits extend
to radii $R_{\rm core}$ between 0.04 and 0.1~pc, but there are several
cores whose BE density distributions extend up to $\sim
0.13$~pc. These values are similar to the values of the BE fits
shown in the literature \citep{Shirley_etal00,
Johnstone_etal00, Alves_etal01, Evans_etal01, Langer_Willacy01,
Harvey_etal01}.  

%numero total de cores rechazados=          30
%numero total de cores con 3 proyecciones BE=          65
%numero total de cores con 2 proyecciones BE=          20
%numero total de cores con 1 projeccion BE          11

\begin{figure} 
\plotone{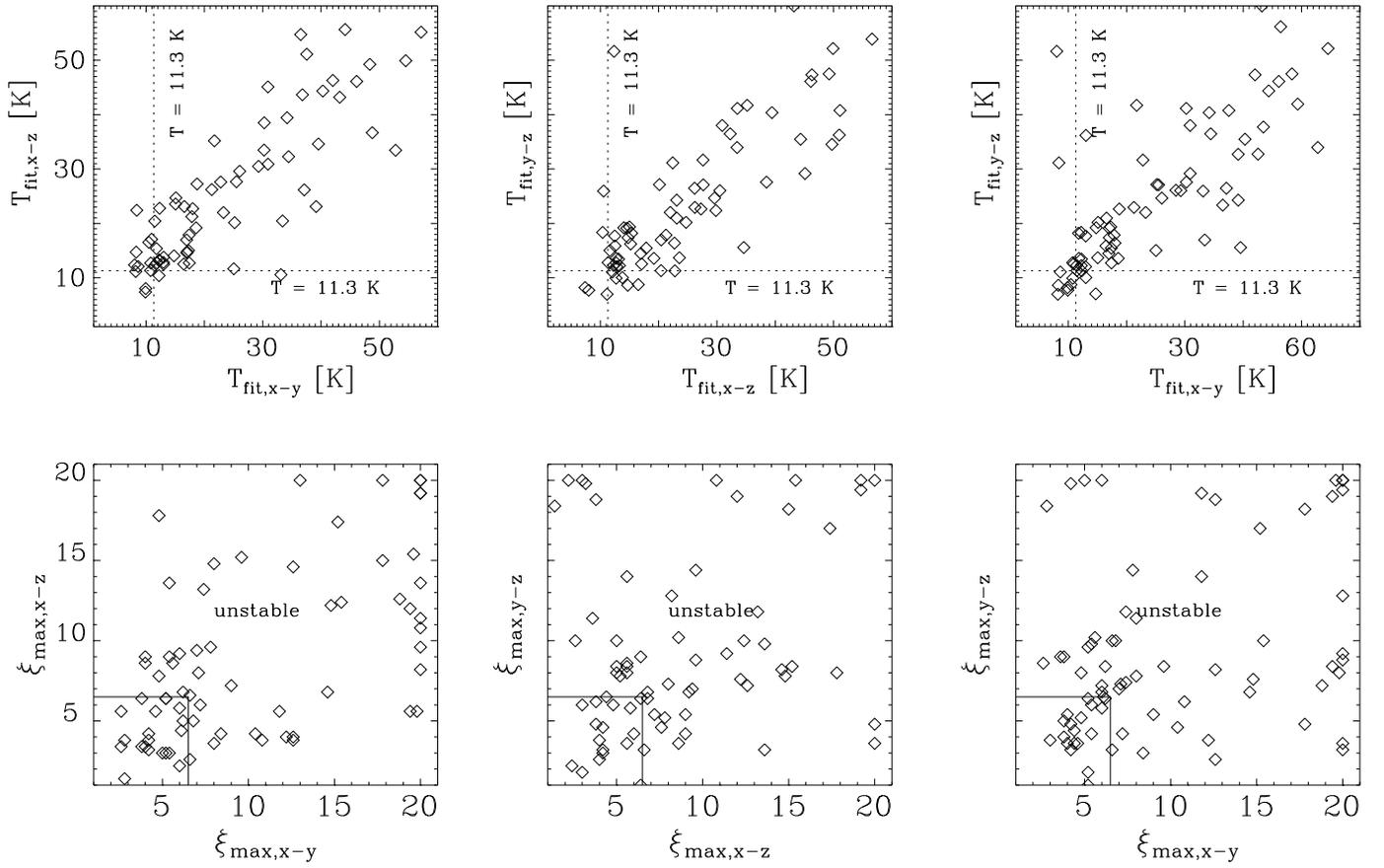}
\caption{Scatter plots of temperature (upper panels),  and \ximax
(lower panels). Left column shows \xz\ against \xy\ 
projections. Middle column shows \yz\ against \xz\ projections, and
right column shows \yz\ against \xy\ projections. Note that if all the
projections had consistent BE profiles, the points should fall along
the identity line in each plot. The square areas in the lower left
corner of lower panels  indicate the regime of nominally stable
cores. 
\label{chi_3D}}
\end{figure}

An important remark is that, amongst the 248 accepted
fits, 195 come from 65 cores that can be fit in all three
directions (\xy, \xz, and \yz) simultaneously. Out of the remaining
fits, 40 come from 20 cores that are fitted in only two of their three
projections, and 11 cores 
resemble BE spheres in only one projection. Furthermore,
cores that are accepted in more than one projection do not generally
yield the same values for \Tfit, and/or \ximax. This fact is
shown in Fig.~\ref{chi_3D}, where we give the values
obtained for the 
temperatures (upper 3 panels) and dimensionless radii of the BE
sphere, \ximax\ (lower 3 panels). Under the light of these results, the
usefulness of BE-type fits to molecular cloud cores is seen to be
suspect. We discuss this issue further in \S\ref{discussion}. 

\subsection{Column density profiles and projection effects}

\begin{figure} 
\begin{center}
\includegraphics[width=0.6\textwidth]{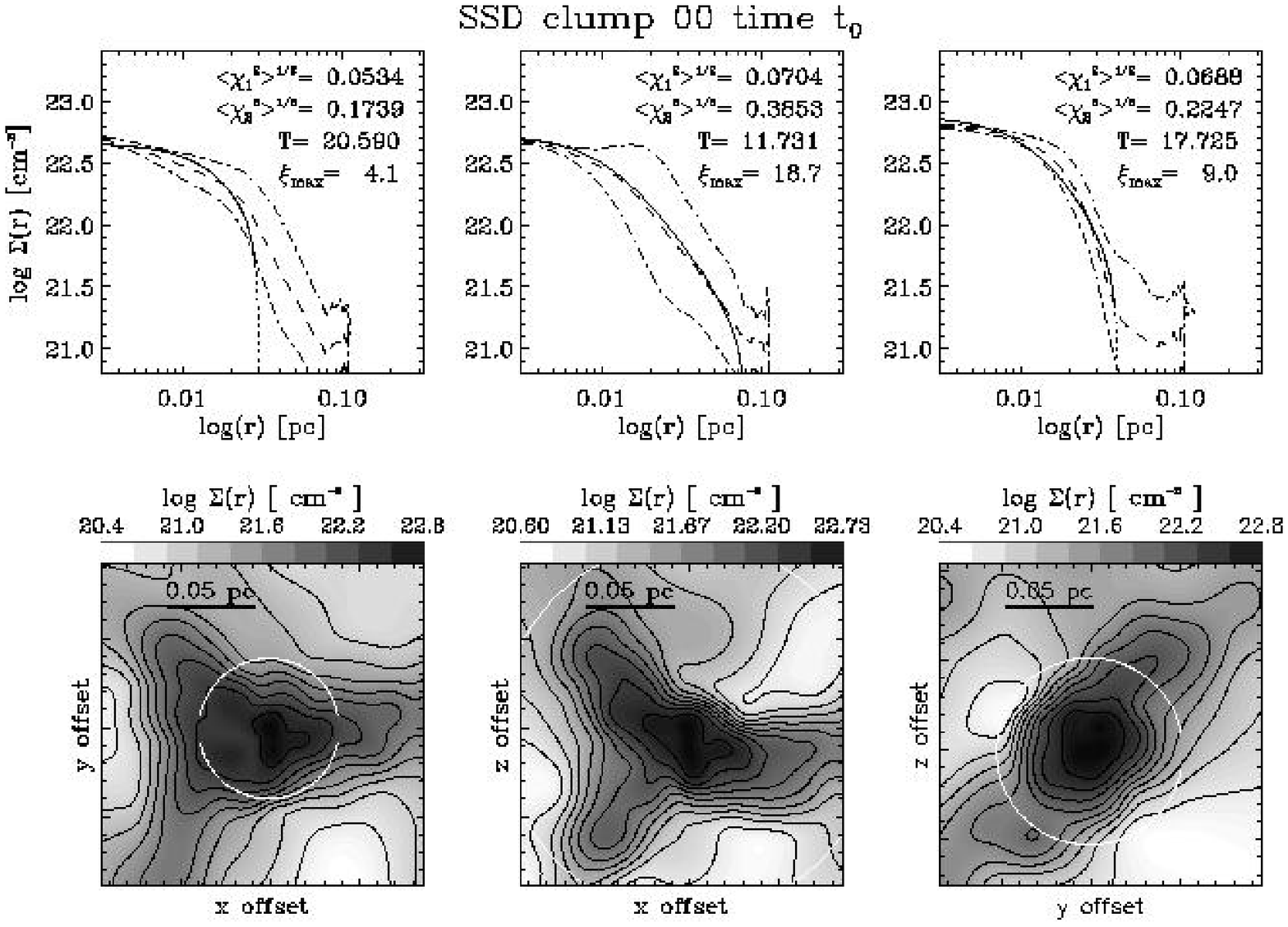}\\[0.3cm]
\includegraphics[width=0.6\textwidth]{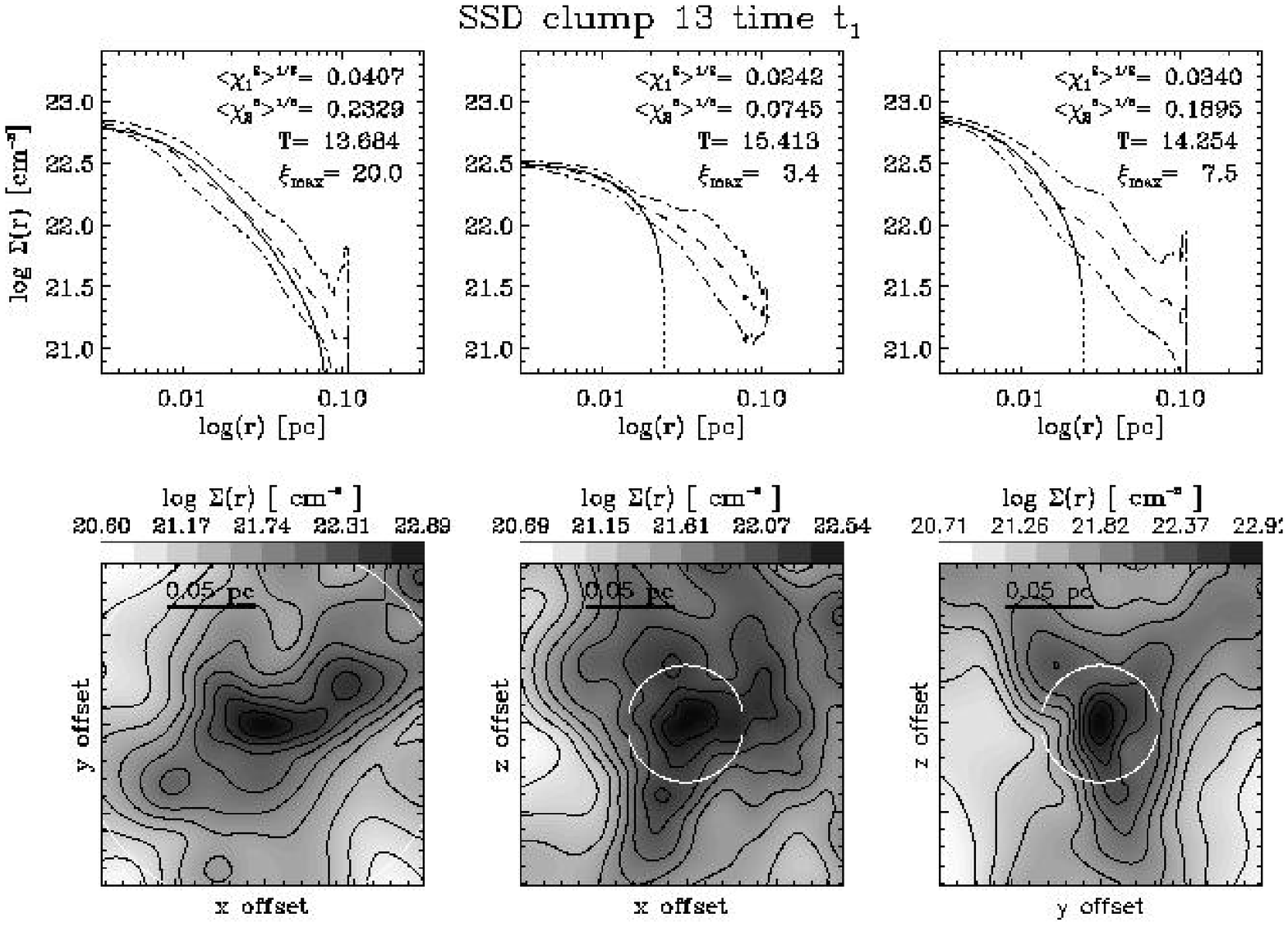} 
\end{center}
\caption{Column density maps and radial profiles for \ssd\ at \tcero\
(a) and \ssd\ at \tuno (b). The white circles show in each case the
size of \Rcore, the radius used in the BE fit. Note the different
morphologies that the same core shows in each projection. 
\label{ssd0}}
\end{figure}

\begin{figure}
\begin{center}
\includegraphics[width=0.6\textwidth]{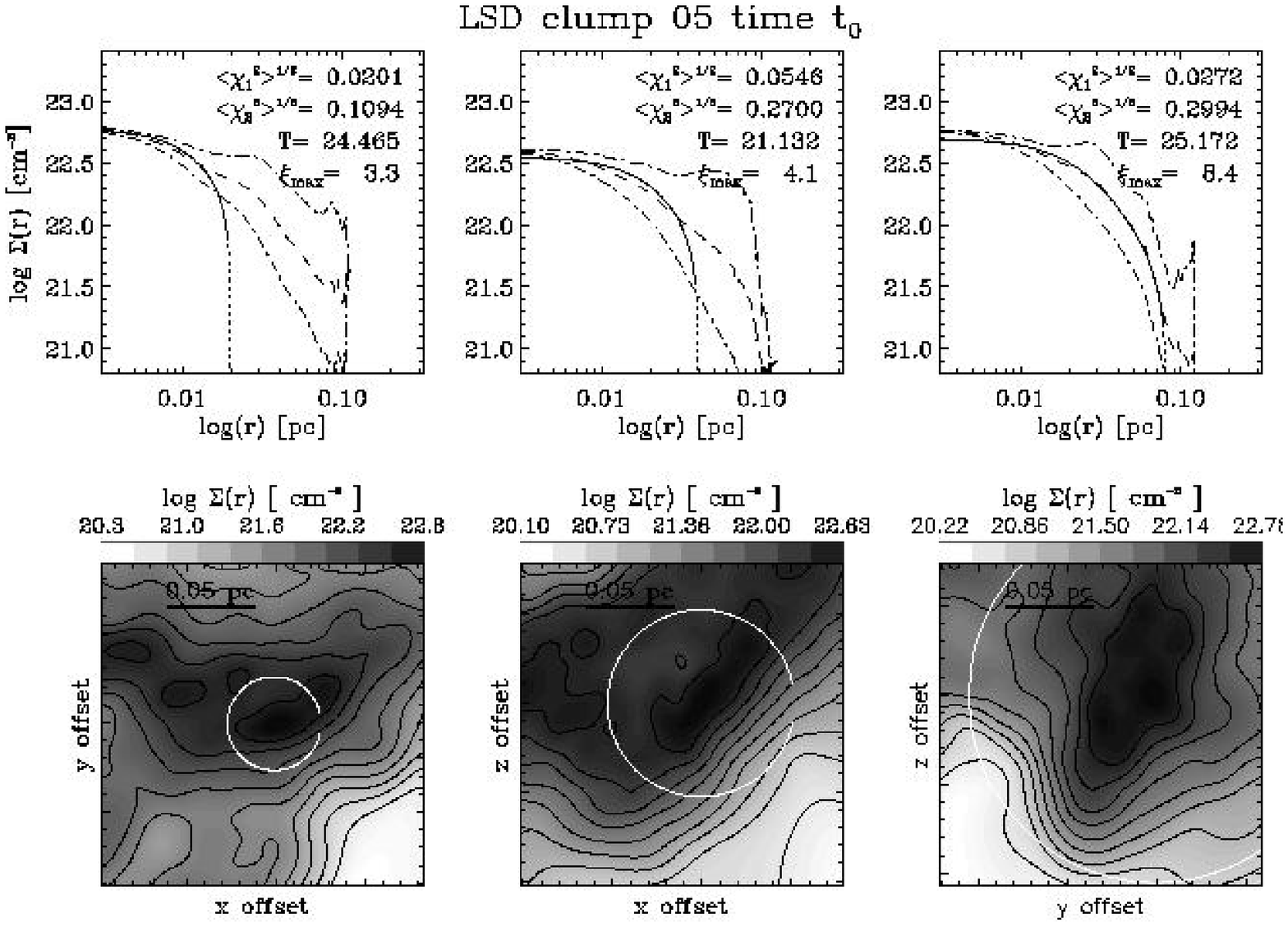}\\[0.3cm]
\includegraphics[width=0.6\textwidth]{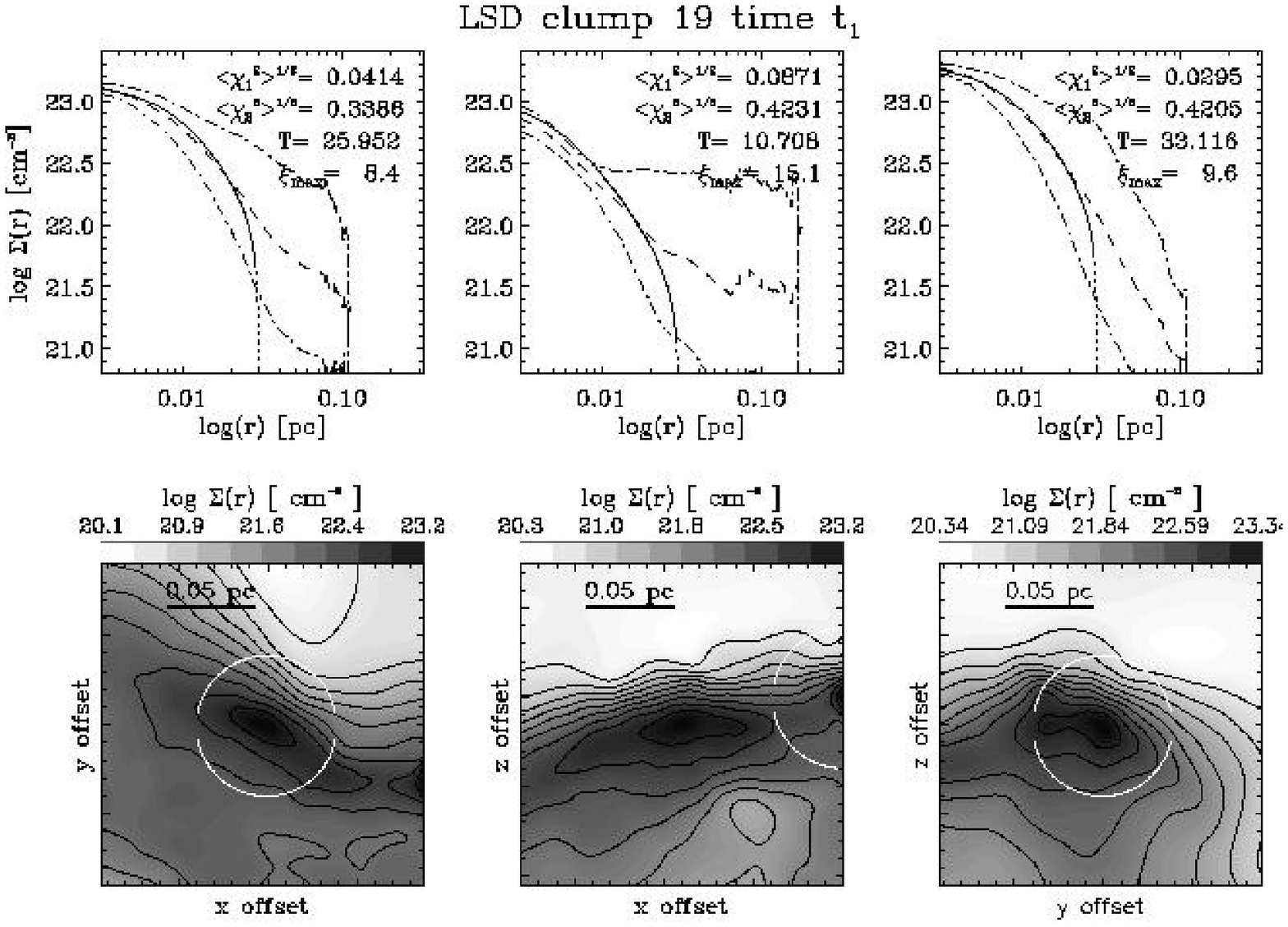} 
\end{center}
%\plotone{f7a.lr.ps}
%\plotone{f7b.lr.ps} 
\caption{Similar to Fig.~\ref{ssd0}, but for \lsd\ at \tcero\ (a) and 
\tuno\ (b).  
\label{lsd0}}
\end{figure}

We now turn to the column density structure of the cores and to the
confusion that may arise due to projection effects. In
Figures~\ref{ssd0} -- \ref{gc0} we show column density maps and radial
profiles of the six cores listed in Table~\ref{tabla:results}. 
%No special criterion was used in choosing these 
%cores, other than the fact that they are representative of each one of the
%physical situations described in \S\ref{simulations} (see also
%Table~\ref{tabla:models}).
%\footnote{For reasons of space, we did not
%include all the plots produced for this analysis. The interested
%reader can download the ps files from {\tt
%http://www.astroscu.unam.mx/\~\ j.ballesteros/BE/ps\_figs.tar.gz}}. 
Upper panels show the logarithmic column density profiles, and lower
panels show the column density maps in logarithmic gray-scale. In the
first, the dashed line corresponds to the actual averaged density 
profile, dotted lines are the average $\pm$ the standard deviation
column density profiles (the ``dispersion'' profiles), and
the solid line denotes the column density of the fitted BE profile. The
left, middle, and right panels respectively depict the \xy, \xz, and
\yz\ projections respectively. The white circle in the lower panels
shows the size of the BE sphere fitted. The first point to notice
is that the column density distribution within these circles is far
from circularly symmetric. Instead, it is highly irregular, often
elongated, and sometimes contains more than one local
maximum. Nevertheless, radial profiles frequently appear soft and
monotonically decreasing. 

In Figs.~\ref{ssd0}a and \ref{ssd0}b we show the maps and profiles for
clump 0 in \ssd\ at \tcero\ and clump 13 in \ssd\ at \tuno,
respectively. For each of the 
three projections of the first case we fit a BE sphere with relatively
good confidence at \Rcore$=$ 0.03, 0.08 and 0.04~pc. We notice that
this core has very
different values of \ximax($=$ 4.1, 18.7, and 9), and of the 
estimated temperature
\Tcore($=$ 20.59, 11.73, and 17.73~K) for each of the projections. In
the case of 
\ssd\ at \tuno, the values obtained using \Rcore$= 0.08$, 0.025, and
0.025 pc are \ximax$=$ 20, 3.4, and 7.5; and $T=$ 13.68, 15.41, and 14.25~K.
%respectively for each projection. From this first two examples we note
%that the same core is fitted with very different values depending on
%the projection under analysis. 
In particular, note that in the case of \ssd\ at \tuno, the three
values for \ximax\ are such that the BE fit will imply unstable,
stable, and unstable (but closer to critical) configurations for the \xy,
\xz, and \yz\ projections, respectively.

%\footnote{It is
%important to mention that these are not necessarily the same core at
%different timesteps. They are just the 11th core found by
%the clumpfinding algorithm at each timestep.}
% another possiblitiy for discussion is ssd core 20 at t1.

Regarding run LSD, we analyze clump  5 at $t=t_0$, and clump 19 at
$t=t_1$. For the former (Fig.~\ref{lsd0}a), we find again
different BE configurations for each projection; in the \xy\ and \xz\
projections we obtain very small values of \ximax (3.3 and 4.1
respectively, implying a core in 
stable equilibrium), and the core is fitted up to only 0.02 and
0.04~pc respectively, while in the \yz\ projection we find
\ximax~=~8.4, and the fit can be made out to 0.08~pc. The fitted
temperatures for this core are 24.47, 21.13, and 25.17 K.
The profiles and the maps for clump 19 in \lsd\ at \tuno\ are shown in
Fig.~\ref{lsd0}b. This core exhibits an elongated structure in each
projection. The fitted values of \ximax~ are 8.4, 15.1, and 9.6,
respectively. The temperatures in each projection are more scattered
than in the previous examples, being $T=25.95$, 10.71, and 33.12.
It is convenient to mention that this core appears in this figure not
because of the great quality of its fit, but to stress the problem of
column density maps. For instance, while the maximum
of the column density is located close to the maximum of the
volumetric density for the \xy\ and \yz\ projections, the maximum
column density of the \xz\ projection is located almost at the edge of
the box. In fact, we have found that frequently, the position of the
maximum column density is shifted by several pixels from the position of the
maximum volumetric density (1 pixel $=$ 0.0012 pc for LSD and SSD, and
0.0014 pc for GC), the \xz\ projection being the most critical case.

Finally, we show cores 4 and 26 for run \gc\ at
\tcero\ and \tuno\ respectively. In this run, cores tend in general
to be  more roundish than in the other cases and thus have
smaller error. Figure \ref{gc0}a  illustrates the
effects of projection. The \xz\ projection (middle panel) shows a
well-defined, round core, and is very well fitted by a BE sphere with
\ximax~$=$~5.1. Nevertheless, this core is just the overlap of two
cores seen in projection, as shown by the \xy\ and \yz\ projections. 
A similar case is presented for GC at $t=t_1$ in Fig.~\ref{gc0}b, where the
\xy\ projection gives an unstable fit (\ximax=8.6), while the other
two projections give stable configurations (5.7 and 2.7 for the \xz\
and \yz, respectively).

Before ending this section, there are some points worth noting. First
of all, our criterion for a ``good'' fit is that 
$\langle \chi_2^2 \rangle^{1/2} > \langle \chi_1^2
\rangle^{1/2}$ (see \S\ref{sec:statistical}). 
The algorithm was constructed to
find the maximum column density in the box, which some times is
shifted from the center of the box. This is because, even if the
maximum volumetric density is well-centered, there is no guarantee
that the column density may will be located at the same place. In
particular, there are some cores that fall at the edge of the box
(as mentioned before, Fig.~\ref{lsd0}b, projection \xz), and others
for which \Rcore\ goes out of the box
boundaries (Fig.'s \ref{ssd0}a projection \xz,
\ref{ssd0}b projection \xy, and \ref{lsd0}a projection \yz). In these
cases, the radial profile was computed by considering only the
density structure inside the box. 

%We have considered to take out of
%the statistics both the cores which fit falls, at least in one point,
%farther than the error curves, and these which maximum radii fall out
%of the boundary of the box. Nevertheless, even thought the statistical
%numbers may change, the main result will still be valid: we can find
%BE-like profiles for cores in a dynamical situation, as can be seen
%from the remaining figures not mentioned in the present paragraph.
%
%Finally, as we mentioned in \S\ref{fitting:sec},
%we perform the fits for values of \ximax\
%between 0.1 and 20. Cores with larger \ximax\ are not taken into
%account, and probably there are some cores that may give a better fit
%for larger \ximax. Nevertheless, the majority of cores show \ximax $<
%6.5$. 

\subsection{The physical conditions in the clumps}\label{sec:nonhydro} 

Throughout this paper, we have repeatedly stated that the cores and
clumps in the simulations are not hydrostatic, but are instead
dynamic, transient entities. This is obviously the case of the cores
selected at \tcero\ in runs LSD and SSD, since self gravity has not
been turned on at this 
stage yet. So, the fact that we have many acceptable fits at this time
is a clear proof that BE fits to the cores' column density profiles do
not provide unambiguous evidence of the cores being in hydrostatic
equilibrium. 

It is moreover instructive to analyze the cores' physical structure, and 
compare it to the observational data. A detailed study of the clump
evolution and a comparison between the kinematics and line profiles of the
simulations and the observations will be
presented in a future contribution. Nevertheless, here we wish to give
just a brief discussion of the 
the density and velocity profiles along the three coordinate
axes for each one of the six cores presented in the previous section, 
in order to show their similarity with observed cores, and further
justify our claim of non-hydrostatic conditions within them. 

\begin{figure} 
\plotone{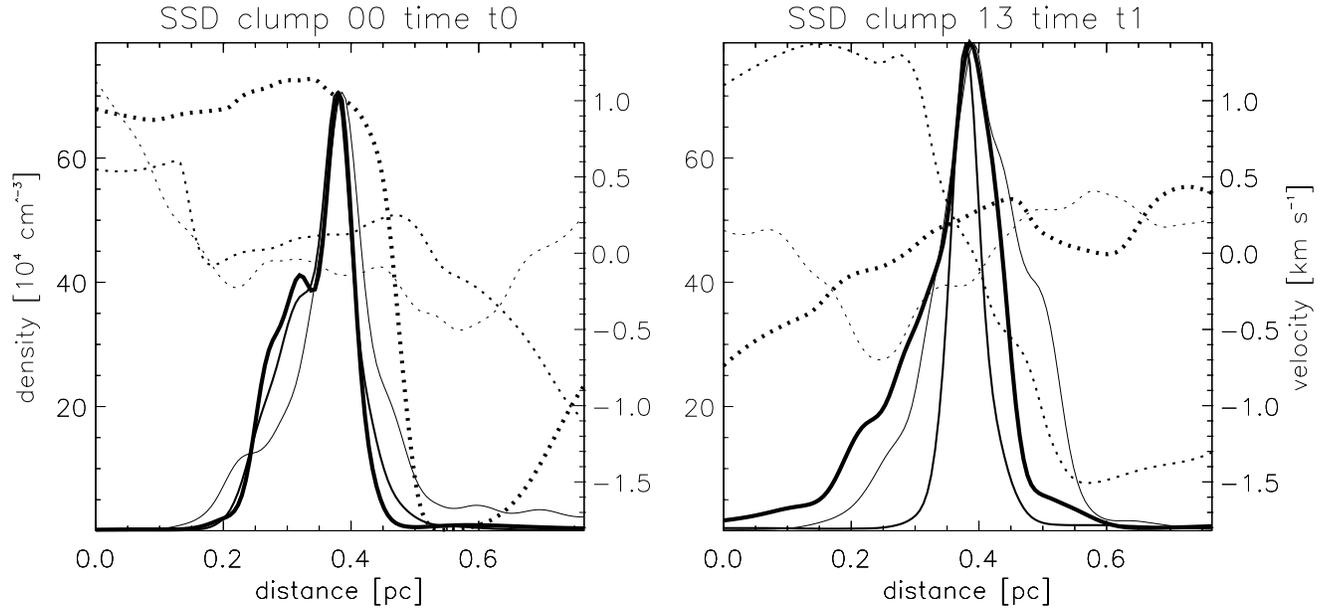}
 \caption{Density (solid lines, left $y$ axis) and velocity (dotted
lines, right $y$ axis) cuts for clumps in Fig.~6 (\ssd), in physical
units (sound speed $c=0.2$ km$\,$s$^{-1}$). Left panel shows the
profiles for clump 0 at \tcero, and right panel shows the profiles for
clump 13 at \tuno. Thin lines represent cuts along the $x$ axis,
intermediate bold lines represent cuts along the $y$ axis, and thick
lines represent cuts along the $z$ axis. See
\S\ref{sec:nonhydro}. Note that the profiles go through the position
of the volumetric density maximum, which does not necessarily
correspond to the position of the column density maximum (center
of the white circle in Fig.~\ref{ssd0}).
\label{ssd_cuts}}
\end{figure}

\begin{figure} 
\plotone{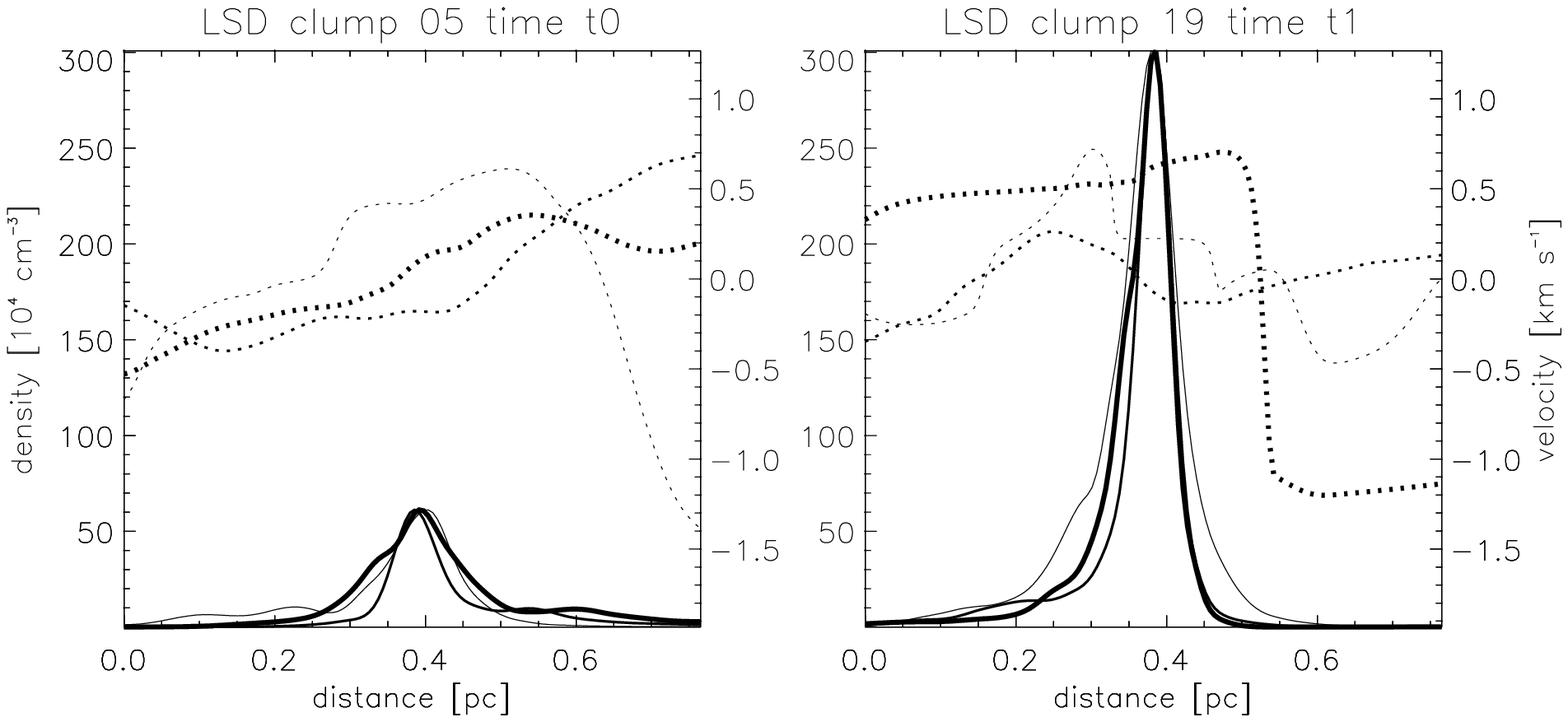}
\caption{Similar to Fig.~\ref{ssd_cuts}, but for  clumps in
Fig.~\ref{lsd0} (\lsd). 
\label{lsd_cuts}}
\end{figure}

Figure~\ref{ssd_cuts} shows density (solid lines) and velocity (dotted
lines) cuts for the clumps
presented in Fig.\ \ref{ssd0} (clumps 0 at \tcero\ and 13 at \tuno); 
Fig.~\ref{lsd_cuts} shows similar cuts for the clumps presented in
Fig.~\ref{lsd0} (clumps 5 at \tcero\ and 19 at
\tuno); and Fig.~\ref{gc_cuts} shows cuts for clumps 4 at \tcero\
and 26 at \tuno, presented in Fig.~\ref{gc0}.
Thin lines denote $x$ axis cuts of the density (solid) and the
$x$-component of the velocity field ($v_x$, dotted). Similarly,
intermediate bold lines denote cuts along the $y$ axis of the density 
and $v_y$, while thick lines denote $z$-axis cuts of the density
and $v_z$. 
%
% chince enrique! aqui estaba esta footnote!
% bueno, creo que quedo mejor en el main text.
%\footnote{Note that in the present analysis we
%are includding cores at \tcero, for which the self-gravity has not
%being turned-on, and can not be in hydrostatic equilibrium.}.
%
From these figures, various points are worth noting. First,
the density profiles are asymmetrical at least in one of the
projections. Second, the velocity profile 
across the clump exhibits, at least in one of the three cuts, a
difference larger than the sound speed (=0.1, in code units). Third,
the velocity gradients exhibited by the cores show that, (a) in the
\ssd\ cores (Fig.~\ref{ssd_cuts}), there are some directions of
contraction (negative gradients of the velocity), and some directions
of expansion (positive gradients). (b) The same occurs for clump 19 at
\tuno\ in  \lsd, but in the case of clump 5 at \tcero\ for \lsd, the
three directions show positive gradients of the velocity field,
suggesting that this clump is 
actually re-expanding. (c) The cores in \gc\ always show negative
gradients, a natural result because these runs are not turbulent, and
gravitational contraction is the only possible mechanism to form
condensations. 

The shapes and amplitudes of the velocity profiles across the clumps
show that they are qualitatively similar to observed cores, in the
following senses: a) The simulated clumps are transonic, exhibiting both
sub- and super-sonic velocity differences across them, similarly to
the reported velocity dispersions for observed cores 
%RSK: citation did not show up correctly in the PS file: (see, e.g.,
%\citet{Hotzel_etal02}). 
 \citep[see, e.g.,][]{Jijina_etal99}. b) The 
velocity patterns include both inflow and outflow, which also occurs in
real cores when observed at high enough resolution
(Myers, Evans \& Ohashi 2000, and references therein).
Thus, our dynamic cores are not dissimilar to observed molecular cloud
cores. Nevertheless, observation of the
evolution of the simulations clearly shows that the cores are not hydrostatic
structures, having typical lifetimes of the order of their crossing
times (Kleesen \& Lin 2003; see also V\'azquez-Semadeni et al.\ 1996).
%They always either re-expand and merge with the
%surrounding medium, or else collapse, depending on the amount of
%gravitational and internal energies. 

\section{Summary and Discussion}\label{discussion}

\subsection{Bonnor-Ebert fits}

We have analyzed cores in SPH simulations of molecular clouds from
%RSK: citation did not work out --  changed
\citet{KBB98},  \citet{Klessen_Burkert00, Klessen_Burkert01} and
\citet{KHM00}. The advantage of using SPH 
simulations over, for example, a regular fixed-grid Eulerian code, is that
in the SPH case, density enhancements are better resolved spatially,
allowing us to make a detailed study of cores at scales between
3\por\diezalamenos 3 pc and 0.3~pc, the relevant scales for molecular
cloud cores.

We have found cores whose angle-averaged column density profiles are well
fitted by BE profiles in a variety of physical situations:
with and without self-gravity (respectively, times \tuno\ and \tcero),
with turbulent driving at large and small
scales (respectively \lsd\ and \ssd), and even cases with merely
random, Gaussian initial density fluctuations (case \gc). Several
results are found. 
First, we have shown that 65\%\ of the column
density profiles studied may resemble BE profiles, in spite of the
fact that the cores are not in hydrostatic equilibrium, and the fact
that at $t=t_0$ the self-gravity has not been included yet for the
turbulent runs \lsd\ and \ssd.

%In the case of the \gc\ runs, the
%core structure is only the result of the initial, arbitrary, random,
%Gaussian density fluctuations. In the case of the \lsd\ and \ssd\
%runs, the core structure is the result of hydrodynamical compressions,
%and must re-expand once the compression gives up, since in an
%isothermal medium there is an excess of thermal pressure at the center
%of the cores.  

Second, we have found that BE fits give temperature values in a range
between 5 and 60~K, with most of them being between 5 and 30~K. The 
fitted central densities range between $\sim$ \diezala 4~cm\alamenos 3
and $\sim$ \diezala 7~cm\alamenos 3, with most of them between
2\por\diezala 5~cm\alamenos 3 and \diezala 6~cm\alamenos 3. These
values are
similar to the values found in the literature \citep{Shirley_etal00,
Johnstone_etal00, Alves_etal01, Evans_etal01, Langer_Willacy01,
Harvey_etal01}. Nevertheless, the fitted values of the temperature and
density in general do not represent the values for
the actual cores. Specifically, the actual
temperature of the cores is always 11.3~K, while 
the fitted value of the density is typically smaller than the actual
density. The values of the non-dimensional radius \ximax\ range
between 2 and 20, with 47\%\ of them below 6.5, and would seem to suggest
that an important fraction of the fitted cores are in hydrostatic 
equilibrium, while in reality they are transient objects.

\subsection{Goodness of fit, and the stability of B68}

A natural question refers to the goodness of our fits. 
We have proposed that two conditions must be
satisfied in order for us to accept a fit.  First, the
core under consideration should not contain a collapsed 
object, i.e.\ it should not yet have formed a
sink particle in our numerical scheme, as we are not able to resolve
such object.  Second, the 
rms separation between the fitted and the angle-averaged
(``mean'') column density profiles must 
be smaller than the rms separation between the dispersion profile and
the mean profile. These two conditions are
met in about 1/2 of the analyzed cores. The typical dispersion curves of the
fits for the accepted cores are not significantly different from the
error bars
of several observational studies \citep{Shirley_etal00,
Johnstone_etal00, Alves_etal01, Evans_etal01, Langer_Willacy01,
Harvey_etal01} and, like in those studies, our fits generally span one
to two orders of magnitude in column density. One notable exception is the
fit of Bok globule B68 by \citet{Alves_etal01}, which has remarkably
small error bars. In that work, the column density map is defined over
$ \sim 1000$ 
positions (the positions of the background stars), which
corresponds to an equivalent resolution of $\sqrt{1000}\sim 30$ pixels
per dimension on the plane of the sky. The reported error bar
at each radius in that paper is calculated as the
standard deviation of the observational uncertainties of all positions
at such radius (J. Alves, private
communication). In fact, those authors comment that their
column density profile is the highest signal-to-noise radial column
density profile ever obtained for Barnard 68. For comparison, in our
simulations we know the value of the column 
density without uncertainty over more than $128^2=16,384$ positions on
the projection plane,
but our rms errors are nevertheless larger, except for some cases in
the \gc\ runs. This appears to suggest
that B68 may actually be an especially smooth and roundish structure,
at least as seen from Earth. Although this is consistent with the fact
that B68 
is actually located within an H~II region, and so the BE paradigm of
thermal pressure confinement by a hotter, more tenuous medium, is
applicable in this case, there are still some pieces of inconsistency
concerning this core:
First, the core is not round,
or elliptical. This makes it implausible that it can be in
precise hydrostatic equilibrium. Second, their recently reported
observations show motions of 0.25$-$0.5 the sound speed, suggesting
that the cloud is near but not precisely in hydrostatic equilibrium.
And third, the fitted Bonnor-Ebert profile by \citet{Alves_etal01}
implies a temperature of 16~K, and, even at this temperature, marginal
instability. The actual temperature is $\sim 10$~K, i.e., a factor of
30-50\% lower that that \citep{Hotzel_etal02}, and thus
the thermal support is even lower. Thus, even within the context of BE
spheres, this core is unstable by a wide margin, and therefore the
hydrostatic model does not appear to provide a satisfactory
explanation of its physical state. In the next
section we suggest a possible alternative for B68.

\subsection{Projection effects}

We have shown that the cores in the simulations are in general
far from spherical, similarly to the situation for actual molecular
cloud cores, and that they
may exhibit substantially different 
column density profiles, depending on the projection direction. Cores
often exhibit different morphologies in different 
directions, and the radius at which they seem to merge with their
surroundings is not unique, also depending on the direction of the
projection. The projection effects play an important role here, and a
core apparently well-defined in one projection may appear as a highly
structured column density
profile in the others. An extreme case is presented
by  \citet{Boss_Hartmann01}, who show that a reasonable Bonnor-Ebert
fit can be obtained even to a disk-like structure seen edge-on.  

\begin{figure} 
\begin{center}
\includegraphics[width=0.6\textwidth]{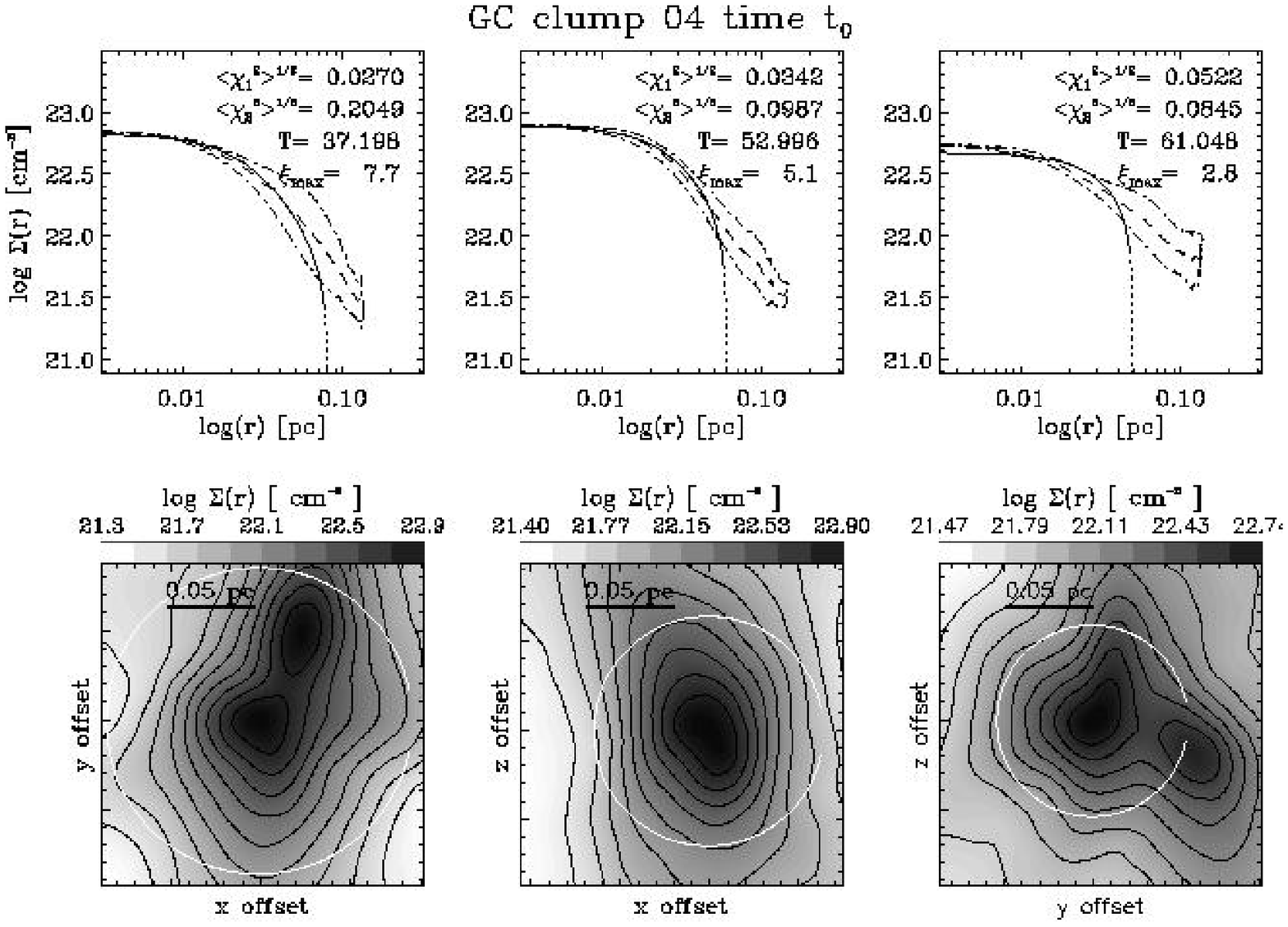}\\[0.3cm]
\includegraphics[width=0.6\textwidth]{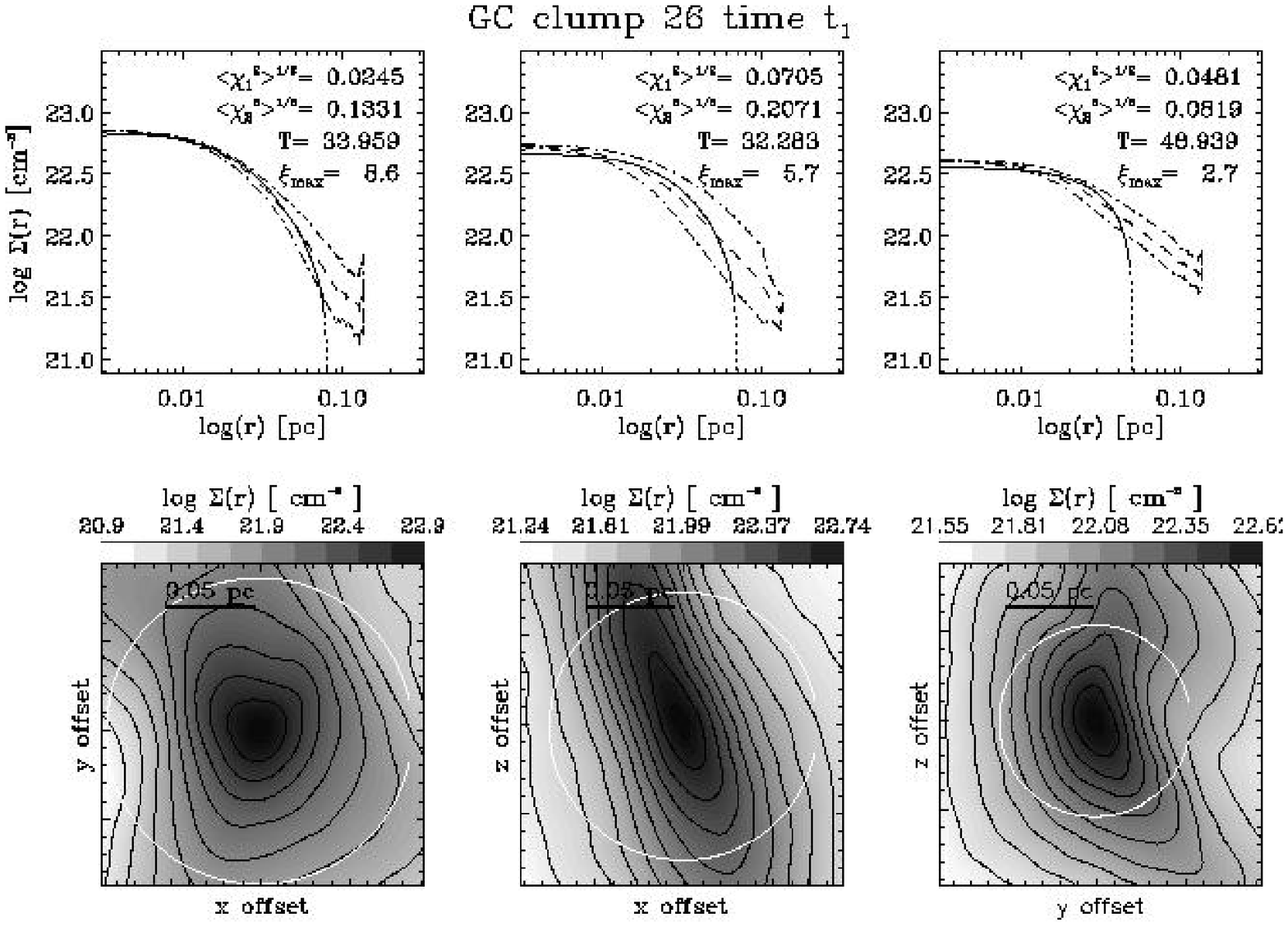} 
\end{center}
%\plotone{f8a.lr.ps}
%\plotone{f8b.lr.ps}
\caption{Similar to Fig.~\ref{ssd0}, but for \gc\ at \tcero\ (a) and 
\tuno\ (b).  
\label{gc0}}
\end{figure}

\begin{figure} 
\plotone{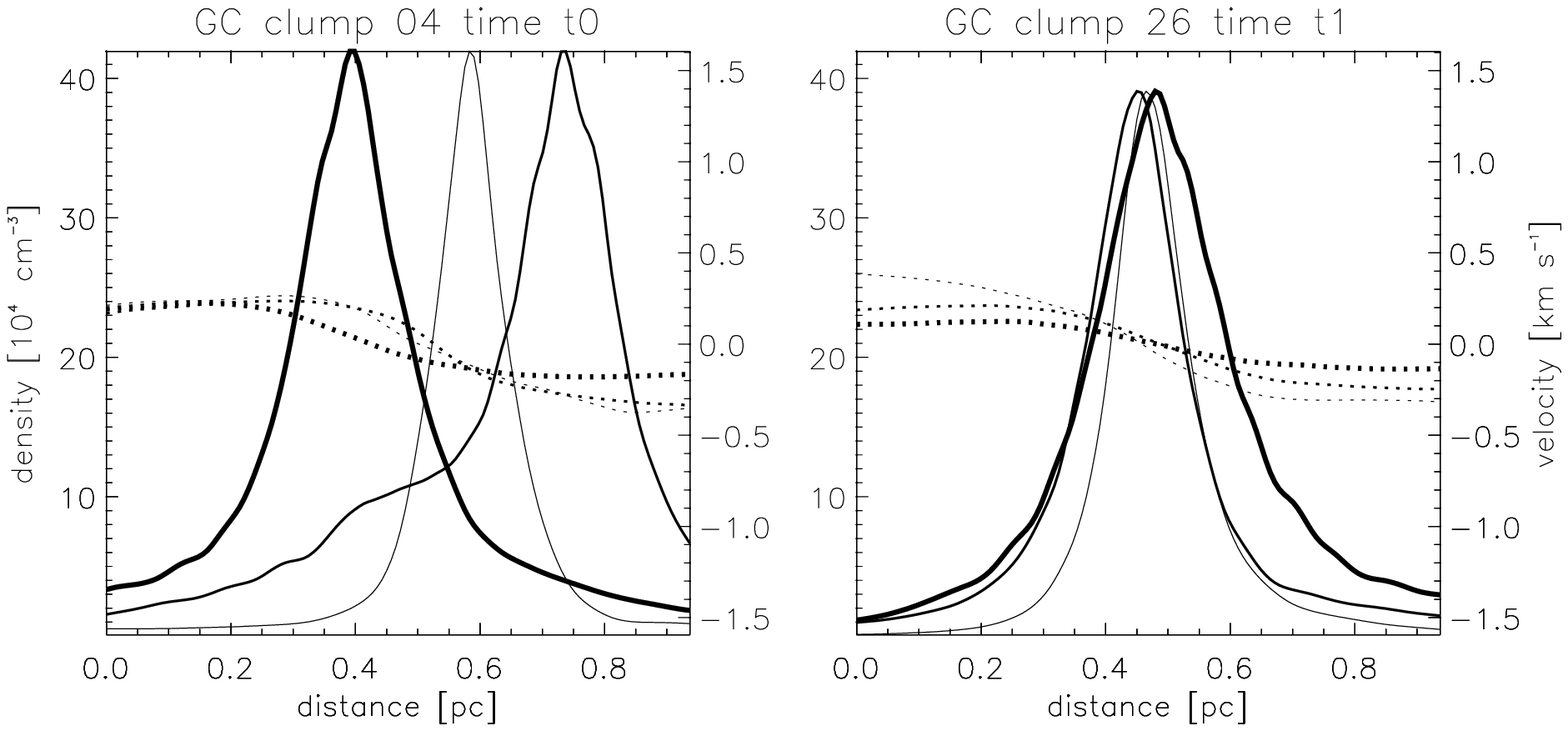}
\caption{Similar to Fig.~\ref{ssd_cuts}, but for  clumps in
Fig.~\ref{gc0} (\gc). 
\label{gc_cuts}}
\end{figure}

In this regard it is important to mention that, in order to minimize
confusion of cores that are in the same line of sight (LOS) but whose
separation is large enough to be considered dynamically disconnected
from each other (as was often the case in the structures analyzed by
Ballesteros-Paredes et al. 1999; Ballesteros-Paredes \& Mac Low 2002),
we have chosen sub-boxes of size one-tenth of the full computational
domain. Thus, the sub-boxes have sizes of 0.154 pc in the \ssd\ and
\lsd\ cases, and of 0.18 pc in the GC case, around the center of the
three-dimensional core, and the LOS integration for producing the
column density maps was performed only over the length of these
sub-boxes.
%
%This implies that, when we constructed the column
%density maps shown here, we did not integrate over the whole
%computational domain, but
%only over scales $\lesssim 0.2$~pc, a scale which has been thought
%that at which the clouds stop being self-similar and/or turbulent
%\citep{Larson95, Goodman_etal98, Barranco_Goodman98}. 
Since, even within
these small length scales there is substantial substructure in the
simulations, it is reasonable to expect that observed regions of
comparable size in molecular clouds should also contain significant
amounts of substructure, the apparent observational smoothness of some
cores possibly being the result of LOS crowding, as occurs for example
in Fig.\ \ref{gc0}.

%dense, small cores may still contain significant sub-structure, even if their
%observational maps appear to be smooth.  

The possibility that there is substructure at such small scales
is consistent with the fact
that most newborn stars belong to multiple systems and thus they must
be born in structured, non-uniform cores. 
Observational evidence that molecular cores at these scales can have
substantial clumpy structure is given by \citet{Velusamy_etal95} and
\citet{Wilner_etal00}, who show that B335, recognized earlier as one
of the best candidates of an isolated, round globule in a collapsing
phase (see Myers, et al. 2000, and references therein),
exhibits a 
complex, asymmetrical structure, and where the physical conditions of
the infalling gas suggest that the standard model of
protostellar collapse fails. 

This possibility, together with inspection of Fig.\ \ref{gc0},
suggests an interesting alternative for the nature of B68. In this
figure, we show one sub-box that has an exceptionally good BE fit
on the \xz\ projection plane, while on the other two projection planes
it is seen to consist of two density peaks. Thus, the
goodness of the BE fit to B68 by \citet{Alves_etal01} is not an
unambiguous proof of the core's BE-like nature, even though, for this
particular core, this is not unlikely either. 
%It is true,
%however, that, B68 being inside an HII region, favors the possibility
%that it is a true BE-like structure. This could not occur in a purely
%isothermal medium (see, e.g., \cite{VSB02}), as BE-structures require
%a hotter, more diffuse confining medium \citet{Ebert55,Bonnor56}. 

%\section{Conclusions}\label{conclusions}

% We have analyzed dense cores in three different physical situations at
% two different timesteps in SPH numerical simulations of the
% interior of molecular clouds. Our analysis has focused on the
% angle-averaged column 
% density profile, to which we can fit Bonnor-Ebert profiles,
% showing that a substantial fraction  (52.9\%) of the cores have
% good BE fits, in spite of the fact that they are either transient or
% collapsing structures. A large number of cores present
% profiles that mimic hydrostatic equilibrium, even if their
% physical situation is not one of equilibrium at all. 
% We have also found that even when the projection is restricted to
% short column lengths, the observed cores are often 
% the result of a structured density field. This result suggests that, if
% the selected models represent
% adequately the interior of molecular clouds, even the densest and
% smallest cores observed and reported in the literature may have
% substantial superposition of structures, regardless the particular
% evolutionary state of each core.

We conclude that the evidence 
%RSK: added
based on the BE fitting procedure
that cores in molecular clouds are
in hydrostatic equilibrium is inconclusive. In order to discriminate
between the standard picture of low-mass star formation that proposes
that cores are quiescent and the turbulent picture that states that
they are dynamical, transient entities, we need more detailed
observations and theoretical work, although we emphasize that, to
date, there is no 
numerical model which allows for a turbulent molecular cloud that has
produced quiescent cores\footnote{Note that some of the simulations by
\citet{Ostriker_etal99} do contain magnetostatic flattened structures
of sizes comparable to the whole computational domain. These
are the result of super-Jeans but sub-critical initial conditions, and
periodic boundary conditions that do not allow for further mass
accretion along field lines. However, we expect that in actual
molecular clouds, cores can always accrete matter, eventually becoming
supercritical, and proceeding to collapse.}, while the cores analyzed
here do show similar physical conditions to those typically reported
for real cores. 

% Thus, we conclude that this
%kind of analysis is not an unambiguous 
%test of the standard model of star formation in which hydrostatic
%structures play a key role.

%Concerning projection effects, we have shown that it is sometimes the case
%that a multi-peak density structure appears as a single object in
%projection.
%
%On the observational point of view, tests of
%the prediction that a large fraction of the observed structure is a
%consequence of superposition effects may be based on taking velocity
%information into account and on observing cores  in
%different chemical tracers. Although this will not be an easy task,
%since cores may show different morphological appearance in different
%molecules, and we in addition expect a fair degree of chemical
%differentiation in the cores \citep{Tafalla_etal02}.

\acknowledgements

We thank L. Hartmann for careful reading of the manuscript.  EVS and
JBP acknowledge support from CONACYT's grants 27752-E and I39318-E
respectively. RSK acknowledges support
by the Emmy Noether Program of the Deutsche Forschungsgemeinschaft
(DFG: KL1358/1) and funding by a NASA astrophysics theory program
supporting the joint Center for Star Formation Studies at NASA-Ames
Research Center, UC Berkeley, and UC Santa Cruz. This research has
made use of NASA's Astrophysics Data System Abstract Service.

%\begin{figure} 
%\plotone{f6a.lr.ps}
%\end{figure}

%\begin{figure} 
%\plotone{BDENSclumP11Z03I09701.no.ps} 
%\caption{\label{ssd1}}
%5\end{figure}

%\begin{figure} 
%\plotone{f7a.lr.ps}
%\end{figure}

%\begin{figure} 
%\plotone{BDENSclumP19Z01I02101.no.ps} 
%\caption{\label{lsd1}}
%\end{figure}

%\begin{figure} 
%\plotone{f8a.lr.ps}
%\end{figure}

%\clearpage

%\end{document}

\begin{deluxetable}{c c c c c c c c c}
\tablecolumns{9}
\tabletypesize{\scriptsize}
\tablecaption{Model properties. \label{tabla:models}}
\tablewidth{0pt}
 \tablehead{
          \colhead{Run}
         &\colhead{Time$^a$}
         &\colhead{Turbulence}
         &\colhead{$\ell_{\rm drv}^b$}
         &\colhead{$M_{\rm rms}^c$}
         &\colhead{$L_{\rm box}^d$}
         &\colhead{$\langle n({\rm H}_2) \rangle^e$}
         &\colhead{$c_{\rm s}^f$}
         &\colhead{Reference$^g$}
 \\%\hline
          \colhead{}
         &\colhead{ [$10^5\,$y] }
         &\colhead{}
         &\colhead{ [pc]}
         &\colhead{ }
         &\colhead{ [pc]}
         &\colhead{ [cm$^{-3}$]}
         &\colhead{ [km$\,$s$^{-1}$]}
         &\colhead{ }
}
 \startdata
         {\rm SSD}
        &{  {\rm 0}}
        &{  {\rm small-scale driven}}
        &{  {\rm 0.19}}
        &{  {\rm 10}}
        &{  {\rm 1.54}}
        &{  {\rm $3.3\times10^3$}}
        &{  {\rm 0.2}}
        &{  ${\cal B}3h$ in KHM00$^h$      }
\\%\hline
         {\rm SSD}
        &{  {\rm 30.6}}
        &{  {\rm small-scale driven}}
        &{  {\rm 0.19}}
        &{  {\rm 10}}
        &{  {\rm 1.54}}
        &{  {\rm $3.3\times10^3$}}
        &{  {\rm 0.2}}
        &{  ${\cal B}3h$ in KHM00$^g$      }
\\         {\rm LSD}
        &{  {\rm 0}}
        &{  {\rm large-scale driven}}
        &{  {\rm 0.77}}
        &{  {\rm 10}}
        &{  {\rm 1.54}}
        &{  {\rm $3.3\times10^3$}}
        &{  {\rm 0.2}}
        &{  ${\cal B}1h$ in KHM00     }
\\%\hline
         {\rm LSD}
        &{  {\rm 2.3}}
        &{  {\rm large-scale driven}}
        &{  {\rm 0.77}}
        &{  {\rm 10}}
        &{  {\rm 1.54}}
        &{  {\rm $3.3\times10^3$}}
        &{  {\rm 0.2}}
        &{  ${\cal B}1h$ in KHM00     }
\\%\hline
         {\rm GC}
        &{  {\rm 3.1}}
        &{  {\rm none}}
        &{  {\rm --- }}
        &{  {\rm --- }}
        &{  {\rm 1.8}}
        &{  {\rm $3.3\times10^3$}}
        &{  {\rm 0.2}}
        &{  ${\cal I}1h$ in KB00     }
\\
         {\rm GC}
        &{  {\rm 6.5}}
        &{  {\rm none}}
        &{  {\rm --- }}
        &{  {\rm --- }}
        &{  {\rm 1.8}}
        &{  {\rm $3.3\times10^3$}}
        &{  {\rm 0.2}}
        &{  ${\cal I}1h$ in KB00     }
\\%\hline
\enddata
\tablenotetext{a}{Time after the onset of self-gravity.} 
\tablenotetext{b}{Scale of energy injection for maintaining a constant
level of turbulence.} 
\tablenotetext{c}{Root mean square Mach number of the turbulent flow.}
\tablenotetext{d}{Total size of the computational box.} 
\tablenotetext{e}{Mean density in the  computational box.}
\tablenotetext{f}{Isothermal sound speed in the simulation.}
\tablenotetext{g}{Corresponding model name in original publication for
further reference: KHM00 -- \citet{KHM00}, KB00 --
\citet{Klessen_Burkert00}.} 
\tablenotetext{h}{${\cal B}3h$ is identical to ${\cal B}3$ but with
$200\,000$ SPH particles. } 
%\tablecomments{\\ tablecomments here, if there are.}
\end{deluxetable}

\clearpage

\begin{deluxetable}{c c c c c c c c c c c}
\tablecolumns{9}
\tabletypesize{\scriptsize}
\tablecaption{Properties of cores. \label{tabla:results}}
\tablewidth{0pt}
\tablehead{
         \colhead{Run}
        &\colhead{Time}
        &\colhead{Core \# }
        &\colhead{Projection}
        &\colhead{$n_{core}$}
        &\colhead{$n_{\rm fit}$}
        &\colhead{$T_{\rm fit}$}
        &\colhead{$\xi_{\rm max}$}
        &\colhead{$R_{\rm core}$}
        &\colhead{$\chi_1^2$}
        &\colhead{$\chi_2^2$}
\\
        \colhead{}
        &\colhead{}
         &\colhead{}
       &\colhead{}
        &\colhead{[cm\alamenos 3]}
        &\colhead{[cm\alamenos 3]}
        &\colhead{K}
        &\colhead{}
        &\colhead{[pc]}
        &\colhead{ }
        &\colhead{ }
}
\startdata 
   {\rm SSD}
   &{\centering {\rm \tcero}}
   &{\centering {\rm 00}}
   &{\centering {\rm x-y}}
   &{\centering {\rm       4.24\por\diezala 5}}
   &{\centering {\rm       7.07\por\diezala 5}}
   &{\centering {\rm       20.59}}
   &{\centering {\rm       4.1}}
   &{\centering {\rm     0.03}}
   &{\centering {\rm     0.053}}
   &{\centering {\rm      0.174}}
   \\  
   {\rm }
   &{\centering {\rm }}
   &{\centering {\rm }}
   &{\centering {\rm x-z}}
   &{\centering {\rm       6.36\por\diezala 5}}
   &{\centering {\rm       7.07\por\diezala 5}}
   &{\centering {\rm       11.73}}
   &{\centering {\rm       18.7}}
   &{\centering {\rm      0.08}}
   &{\centering {\rm      0.070}}
   &{\centering {\rm      0.385}}
   \\  
  {\rm }
   &{\centering {\rm }}
   &{\centering {\rm }}
   &{\centering {\rm y-z}}
   &{\centering {\rm       7.07\por\diezala 5}}
   &{\centering {\rm       7.07\por\diezala 5}}
   &{\centering {\rm       17.73}}
   &{\centering {\rm       9.0}}
   &{\centering {\rm      0.04}}
   &{\centering {\rm      0.069}}
   &{\centering {\rm      0.225}}
   \\  
   {\rm SSD}
   &{\centering {\rm \tuno}}
   &{\centering {\rm 13}}
   &{\centering {\rm x-y}}
   &{\centering {\rm       9.43\por\diezala 5}}
   &{\centering {\rm       7.86\por\diezala 5}}
   &{\centering {\rm       13.68}}
   &{\centering {\rm       20.0}}
   &{\centering {\rm     0.08}}
   &{\centering {\rm     0.041}}
   &{\centering {\rm      0.233}}
   \\  
   {\rm }
   &{\centering {\rm }}
   &{\centering {\rm }}
   &{\centering {\rm x-z}}
   &{\centering {\rm       3.14\por\diezala 5}}
   &{\centering {\rm       7.86\por\diezala 5}}
   &{\centering {\rm       15.41}}
   &{\centering {\rm       3.4}}
   &{\centering {\rm     0.025}}
   &{\centering {\rm     0.024}}
   &{\centering {\rm     0.07}}
   \\  
   {\rm }
   &{\centering {\rm }}
   &{\centering {\rm }}
   &{\centering {\rm y-z}}
   &{\centering {\rm       1.49\por\diezala 6}}
   &{\centering {\rm       7.86\por\diezala 5}}
   &{\centering {\rm       14.25}}
   &{\centering {\rm       7.5}}
   &{\centering {\rm     0.025}}
   &{\centering {\rm     0.035}}
   &{\centering {\rm      0.190}}
   \\  
   {\rm LSD}
   &{\centering {\rm \tcero}}
   &{\centering {\rm 05}}
   &{\centering {\rm x-y}}
   &{\centering {\rm       6.73\por\diezala 5}}
   &{\centering {\rm       6.12\por\diezala 5}}
   &{\centering {\rm       24.47}}
   &{\centering {\rm       3.3}}
   &{\centering {\rm     0.02}}
   &{\centering {\rm     0.020}}
   &{\centering {\rm      0.109}}
   \\  
   {\rm }
   &{\centering {\rm }}
   &{\centering {\rm }}
   &{\centering {\rm x-z}}
   &{\centering {\rm       1.84\por\diezala 5}}
   &{\centering {\rm       6.12\por\diezala 5}}
   &{\centering {\rm       21.13}}
   &{\centering {\rm       4.1}}
   &{\centering {\rm      0.04}}
   &{\centering {\rm     0.055}}
   &{\centering {\rm      0.270}}
   \\  
  {\rm }
   &{\centering {\rm }}
   &{\centering {\rm }}
   &{\centering {\rm y-z}}
   &{\centering {\rm       3.06\por\diezala 5}}
   &{\centering {\rm       6.12\por\diezala 5}}
   &{\centering {\rm       25.17}}
   &{\centering {\rm       8.4}}
   &{\centering {\rm      0.08}}
   &{\centering {\rm     0.027}}
   &{\centering {\rm      0.300}}
   \\  
   {\rm LSD}
   &{\centering {\rm \tuno}}
   &{\centering {\rm 19}}
   &{\centering {\rm x-y}}
   &{\centering {\rm       2.24\por\diezala 6}}
   &{\centering {\rm       3.74\por\diezala 6}}
   &{\centering {\rm       25.95}}
   &{\centering {\rm       8.4}}
   &{\centering {\rm     0.03}}
   &{\centering {\rm     0.041}}
   &{\centering {\rm      0.339}}
   \\  
   {\rm }
   &{\centering {\rm }}
   &{\centering {\rm }}
   &{\centering {\rm x-z}}
   &{\centering {\rm       2.99\por\diezala 6}}
   &{\centering {\rm       3.74\por\diezala 6}}
   &{\centering {\rm       10.71}}
   &{\centering {\rm       15.1}}
   &{\centering {\rm     0.03}}
   &{\centering {\rm     0.087}}
   &{\centering {\rm     0.423}}
   \\  
   {\rm }
   &{\centering {\rm }}
   &{\centering {\rm }}
   &{\centering {\rm y-z}}
   &{\centering {\rm       3.74\por\diezala 6}}
   &{\centering {\rm       3.74\por\diezala 6}}
   &{\centering {\rm       33.12}}
   &{\centering {\rm       9.6}}
   &{\centering {\rm     0.03}}
   &{\centering {\rm     0.030}}
   &{\centering {\rm      0.421}}
   \\ 
   {\rm GC}
   &{\centering {\rm \tcero}}
   &{\centering {\rm 04}}
   &{\centering {\rm x-y}}
   &{\centering {\rm       3.77\por\diezala 5}}
   &{\centering {\rm       4.19\por\diezala 5}}
   &{\centering {\rm       37.20}}
   &{\centering {\rm       7.7}}
   &{\centering {\rm     0.08}}
   &{\centering {\rm     0.027}}
   &{\centering {\rm      0.205}}
   \\  
   {\rm }
   &{\centering {\rm }}
   &{\centering {\rm }}
   &{\centering {\rm x-z}}
   &{\centering {\rm       4.19\por\diezala 5}}
   &{\centering {\rm       4.19\por\diezala 5}}
   &{\centering {\rm       53.0}}
   &{\centering {\rm       5.1}}
   &{\centering {\rm     0.06}}
   &{\centering {\rm     0.034}}
   &{\centering {\rm     0.099}}
   \\  
  {\rm }
   &{\centering {\rm }}
   &{\centering {\rm }}
   &{\centering {\rm y-z}}
   &{\centering {\rm       2.10\por\diezala 5}}
   &{\centering {\rm       4.19\por\diezala 5}}
   &{\centering {\rm       61.05}}
   &{\centering {\rm       2.8}}
   &{\centering {\rm     0.05}}
   &{\centering {\rm     0.052}}
   &{\centering {\rm     0.085}}
   \\  
   {\rm GC}
   &{\centering {\rm \tuno}}
   &{\centering {\rm 26}}
   &{\centering {\rm x-y}}
   &{\centering {\rm       4.30\por\diezala 5}}
   &{\centering {\rm       3.91\por\diezala 5}}
   &{\centering {\rm       33.96}}
   &{\centering {\rm       8.6}}
   &{\centering {\rm     0.08}}
   &{\centering {\rm     0.025}}
   &{\centering {\rm      0.133}}
   \\     
   {\rm }
   &{\centering {\rm }}
   &{\centering {\rm }}
   &{\centering {\rm x-z}}
   &{\centering {\rm       2.34\por\diezala 5}}
   &{\centering {\rm       3.91\por\diezala 5}}
   &{\centering {\rm       32.28}}
   &{\centering {\rm       5.7}}
   &{\centering {\rm     0.07}}
   &{\centering {\rm      0.071}}
   &{\centering {\rm      0.207}}
   \\  
   {\rm }
   &{\centering {\rm }}
   &{\centering {\rm }}
   &{\centering {\rm y-z}}
   &{\centering {\rm       1.56\por\diezala 5}}
   &{\centering {\rm       3.91\por\diezala 5}}
   &{\centering {\rm       48.94}}
   &{\centering {\rm       2.7}}
   &{\centering {\rm     0.05}}
   &{\centering {\rm     0.048}}
   &{\centering {\rm     0.082}}
\\%\hline
\enddata
%\tablenotetext{1}{Note that this projection is not used for
%statistics, since \xiunorms $>$ \xidosrms}
%\tablecomments{comentario}
\end{deluxetable}

\end{document}